\documentclass[aps,prb,twocolumn,showpacs,preprintnumbers,amsmath,amssymb,superscriptaddress]{revtex4-2}

\usepackage{graphicx}
\usepackage{dcolumn}
\usepackage{bm}
\usepackage{hyperref}
\hypersetup{
	colorlinks,
	citecolor=red,
	filecolor=black,
	linkcolor=blue,
	urlcolor=blue
}
\usepackage{lipsum}
\usepackage{natbib}

\usepackage{mathptmx}
\usepackage{dsfont}

\usepackage{graphicx}
\usepackage{dcolumn}

\usepackage[utf8]{inputenc}
\usepackage{calc}
\usepackage{accents}

\newcommand{\nc}{\newcommand}
\nc{\be}{\begin{equation}}
\nc{\ee}{\end{equation}}
\nc{\bea}{\begin{eqnarray}}
\nc{\eea}{\end{eqnarray}}
\nc{\bean}{\begin{eqnarray*}}
\newcommand{\derd}{\, \mathrm d}
\nc{\eean}{\end{eqnarray*}}
\nc{\mb}{\mbox}
\nc{\rnc}{\renewcommand}
\nc{\vk}{\mb{\bf k}}
\nc{\vp}{\mb{\bf p}}
\nc{\vn}{\mb{\bf n}}
\nc{\vq}{\mb{\bf q}}
\nc{\rr}{\mb{\bf r}}
\nc{\vz}{\hat {\mb{\bf z}}}
\nc{\vj}{\mb{\boldmath$j$}}
\nc{\vg}{\mb{\boldmath$g$}}
\nc{\x}{\mb{\boldmath$x$}}
\nc{\A}{\mb{\boldmath$A$}}
\nc{\va}{\mb{\boldmath$a$}}
\nc{\vs}{\mb{\boldmath$\sigma$}}
\nc{\vpi}{\mb{\boldmath$\pi$}}
\nc{\nab}{\nabla}
\nc{\X}{\sf x}
\nc{\kk}{{\bf k}}
\nc{\pp}{{\bf p}}
\nc{\uspin}{{\uparrow}}
\nc{\dspin}{{\downarrow}}
\nc{\vecq}{{\bf q}}
\nc{\veck}{{\bf k}}
\nc{\vecp}{{\bf p}}
\nc{\vecl}{{\bf l}}
\nc{\vecr}{{\bf r}}
\nc{\vecx}{{\bf x}}
\nc{\vecR}{{\bf R}}
\nc{\vecG}{{\bf G}}
\nc{\vecA}{{\bf A}}
\nc{\vecpi}{{\bf \pi}}
\nc{\vecL}{{\bf L}}
\nc{\vecK}{{\bf K}}

\nc{\argg}{\text{Arg}}
\nc{\bd}{\textbf}
\nc{\bds}{\boldsymbol}
\nc{\ham}{\hat{\mathcal{H}}}
\nc{\im}{\text{Im}}
\nc{\la}{\langle}
\nc{\ra}{\rangle}
\nc{\re}{\text{Re}}
\nc{\rn}[1]{%
	\textup{\uppercase\expandafter{\romannumeral#1}}%
}
\nc{\sgn}{\text{Sgn}}
\nc{\tit}{\textit}
\nc{\tr}{\text{Tr}}

\nc{\les}{\leqslant}
\nc{\ges}{\geqslant}
\makeatletter
\nc*{\rom}[1]{\expandafter\@slowromancap\romannumeral #1@}

\def\be{\begin{eqnarray}}
\def\ee{\end{eqnarray}}

\newtheorem{thm}{Theorem}[section]
\newtheorem{propt}[thm]{Proposition}

\begin{document}

\title{Geometric superfluid weight of composite bands in multiorbital superconductors}
\author{Guodong Jiang}
\affiliation{Department of Physics, University of Nevada, Reno, Reno NV 89502, USA}
\affiliation{Department of Applied Physics, Aalto University School of Science, FI-00076 Aalto, Finland}
\author{Yafis Barlas}
\affiliation{Department of Physics, University of Nevada, Reno, Reno NV 89502, USA}

\begin{abstract}
The superfluid weight of an isolated flat band in multiorbital superconductors contains contributions from the band's quantum metric and a lattice geometric term that depends on the orbital positions in the lattice. Since the superfluid weight is a measure of the superconductor's energy fluctuation, it is independent of the lattice geometry, leading to the minimal quantum metric of a band [Phys. Rev. B 106, 014518 (2022)]. Here, a perturbation approach is developed to study the superfluid weight and its lattice geometric dependence for composite bands. When all orbitals exhibit uniform pairing, the quantum geometric term contains each band's contribution and an inter-band contribution between every pair of bands in the composite. Based on a band representation analysis, they provide a topological lower bound for the superfluid weight of an isolated composite of flat bands. Using this perturbation approach, an analytical expression of the lattice geometric contribution is obtained. It is expressed in terms of Bloch functions, providing a convenient formula to calculate the superfluid weight for multiorbital superconductors.
\end{abstract}

\maketitle

\section{Introduction}
The discovery of superconductivity in twisted two-dimensional (2D) crystals \cite{Cao2018,Yankowitz1059,Cao2018second,Chen2019,Lu2019,Sharpe605,PhysRevLett.123.197702,CStDLG_2020,CStDLG_Pablo,CStDLG_Kim,Polshyn_2020,CherntMLG_Cory,SCMAtTLGPablo,TripletSCtTLGPablo} has brought focus to the nature of superconductivity in flat or weakly dispersive bands~\cite{PhysRevLett.111.046604,peotta2015superfluidity,julku2016geometric,tovmasyan2016effective,liang2017band,torma2018quantum,iskin2018exposing,hu2019geometric,xie2020topology,PhysRevB.102.184504,hofmann2020,julku2020superfluid,julku2021quantum,PhysRevLett.127.246403,torma2022superconductivity,chan2022pairing,huhtinen2022revisiting,herzog2022manybody}. While the pairing mechanism remains unresolved, the superconducting fluctuations in flat bands are exotic enough to warrant a complete understanding ~\cite{PhysRevB.98.220504,PhysRevX.8.031089,PhysRevLett.121.087001,PhysRevB.98.075154,PhysRevB.98.121406,PhysRevLett.121.217001,PhysRevX.8.041041,PhysRevLett.122.026801,PhysRevB.100.205113,PhysRevLett.127.217001,PhysRevB.104.L121116,PhysRevLett.121.257001,PhysRevB.99.165112,PhysRevLett.122.257002,PhysRevB.98.241412,PhysRevB.98.054515,GuineaPNAS,SkyrmionSC}. In particular, is the question of superfluid weight (SW) in flat band superconductors, which serves as an essential criteria to distinguish superconductors from metals~\cite{scalapino1992superfluid,scalapino1993insulator}. The SW in conventional superconductors, irrespective of the symmetry of order parameter, is inversely proportional to the effective mass~\cite{schrieffer1999theory,Leggett1998}, ruling out flat-band superconductivity. However, in multiorbital superconductors, band geometry rescues superconductivity by providing a finite SW, which is proportional to the quantum metric of the band \cite{peotta2015superfluidity,julku2016geometric,tovmasyan2016effective,liang2017band,iskin2018exposing,torma2018quantum,hu2019geometric,xie2020topology,PhysRevB.102.184504,hofmann2020,julku2021quantum,PhysRevLett.127.246403,torma2022superconductivity}. This quantum geometric SW dominates in flat band superconductors like twisted 2D crystals \cite{hu2019geometric,xie2020topology,Tian2023}. Moreover, geometric effects can result in transitions to exotic superconducting states~\cite{PhysRevB.106.184507,chen2023,jiang2023}, which necessitates the need for further understanding the role of band geometry in multiorbital superconductors.

Up to now, most studies of the geometric SW have been restricted to a single isolated band \cite{peotta2015superfluidity,tovmasyan2016effective,liang2017band,iskin2018exposing,torma2018quantum,hu2019geometric,xie2020topology,PhysRevB.102.184504,hofmann2020,julku2021quantum,PhysRevLett.127.246403,torma2022superconductivity}. This is valid when the pairing potential is comparable with the bandwidth of the active band near the Fermi level but much smaller than the bandgap to other remote bands. However, many examples exist where multiple bands lie within or close to the pairing interaction window, violating this single-band projection. In these cases, the inter-band geometric effects \cite{julku2016geometric,julku2020superfluid,iskin2022effective,mera2022nontrivial} can be significant. In addition, quantum geometric quantities like the Berry curvature and quantum metric are usually large near the local $\bd{k}$-points where the remote bands are energetically close to the active band. Therefore, it is essential to study the geometric SW when a composite of bands is within or near the interaction range.

In this paper, we employ a band projection formalism to study the SW of composite bands. This projection to the composite transforms the orbital basis to band basis but projects to a few bands most relevant to the interaction energy scale determined by the pairing Hamiltonian. We then develop a perturbation approach to calculate the SW, which is similar to the $\bd{k}\cdot\bd{p}$ method of calculating the effective mass tensor of bands in solids. This approach provides an alternative derivation of the general results in Ref.~\cite{peotta2015superfluidity,liang2017band} and identifies each term of the quantum geometric SW from transitions between Bogoliubov bands. The quantum geometric SW of composite bands contains each band's contribution and an inter-band contribution between every pair of bands in the composite. This separation of SW emphasizes the role of inter-band quantum metric, which originates from the derivative of inter-band pairing functions. Although the inter-band geometric terms are negative semidefinite and reduce the SW, the total SW is always positive semidefinite, implying the local stability of the zero center-of-mass momentum (CMM) state in the uniform pairing channel.

Analyzing the reduction of superfluid weight due to the inter-band pairing, we derive a topological lower bound for the SW of an isolated composite of flat bands (ICFB), Eq. \ref{eq:improvebound}. The necessity of such a lower bound arises from a very fundamental question, as in realistic materials, the scenario of a single isolated flat band is rare. Instead, many flat or weakly dispersive bands can coexist in the interaction window. In the simplest case, one can ask if there is a topological lower bound for the SW when two flat bands of opposite Chern numbers $\pm1$ are both within the pairing interaction window.

An analysis of this case from the perspective of band representation (BR) turns out to be illuminating. If the two bands are constructed from the same set of Wannier orbital basis of an elementary band representation (EBR) \cite{bradlyn2017,zak1980symmetry}, their topological lower bounds tend to cancel, $|+1+(-1)|=0$. In contrast, when they belong to different EBRs, their topological lower bounds will add, $|+1|+|-1|=2$. The latter is due to a suppression of the reduction effect by inter-band pairings. This hints that a larger SW can be achieved if a multiorbital superconductor is doped to a set of bands that contain multiple incomplete EBRs.

In the second half of this paper, we study the relation between lattice geometry and SW for multiorbital superconductors. Recently, it was shown that an additional contribution encoding the modification of order parameters with orbital positions in a unit cell must be included in the calculation of SW \cite{chan2022pairing,huhtinen2022revisiting}. This lattice geometric term, combined with the quantum geometric term, makes the total SW invariant with the choice of orbital positions \cite{huhtinen2022revisiting}, therefore lattice geometry-independent \cite{simon2020}. Using the perturbation method, we derive an analytical expression of the lattice geometric term in terms of Bloch functions, Eq. \eqref{eq:ds2final}. This expression provides a convenient way of computing the lattice geometric term numerically without solving gap equations.

The rest of this paper is organized as follows. In Sec. \ref{sec:swcomp}, we develop the projection procedure for composite bands and present the quantum geometric SW calculated from the perturbation method. Sec.~\ref{sec:topobound} contains an analysis of the inter-band effect on the geometric SW and establishes a topological lower bound for an ICFB. Sec. \ref{sec:mqm} studies the derivative of gap equation with respect to CMM using the perturbation method, thereby obtaining the second derivative of grand potential with order parameters and analyzing its properties. In Sec. \ref{sec:newformula}, we show the geometry independence of the SW and provide a convenient formula for computing the lattice geometric term. Finally, we conclude our findings in Sec. \ref{sec:discussion}.
\section{Superfluid weight of Composite bands }
\label{sec:swcomp}
We start from the tight-binding lattice Hamiltonian with intra-orbital attractive interactions ($U_\alpha>0$):
\begin{align}
\label{eq:modelh}
\ham= \sum_{ij,\alpha\beta,\sigma} t^{\sigma}_{ij,\alpha\beta} c^\dagger_{i\alpha\sigma} c_{j\beta\sigma} - \sum_{i,\alpha}U_{\alpha} \hat{n}_{i\alpha\uparrow} \hat{n}_{i\alpha\downarrow},
\end{align}
where $i,j$ label the unit cell, $\alpha,\beta = 1,..., s$ label the orbitals in a unit cell, and $\sigma=\uparrow,\downarrow$ labels the spin. The tight-binding matrix elements, $t^{\sigma}_{ij,\alpha\beta}$ encode the hopping pattern and amplitudes from orbital $(j,\beta)$ to $(i,\alpha)$. This hopping graph determines the band dispersion and topology. The second term is the onsite density-density interaction, with $\hat{n}_{i\alpha\sigma}=c^\dagger_{i\alpha\sigma}c_{i\alpha\sigma}$ the number operator. We assume time-reversal symmetry (TRS) with zero spin-orbit coupling, thereby restricting to singlet pairing.

For attractive pairing interactions, we assume a BCS-type ground state and perform the usual mean-field decoupling of Eq.~\eqref{eq:modelh} to arrive at the mean-field Hamiltonian $\ham_{MF}$. Throughout this paper, we focus on an orbital-independent pairing matrix $\hat{\Delta}=\Delta\hat{\mathcal{I}}_s$, referred to as the uniform pairing condition (UPC) in literature \cite{peotta2015superfluidity,julku2016geometric,liang2017band}. Here $\hat{\mathcal{I}}_s$ is the $s$-dim identity matrix in orbital space. This choice of pairing channel can always be made for any intra-orbital interactions obeying TRS by tuning parameters $U_\alpha$. It corresponds to a particular solution channel to the BCS gap equation, which always leads to a positive semidefinite SW tensor.

The superconducting fluctuations to finite CMM can be described within the same framework by a reduced mean-field Hamiltonian (see App. \ref{app:finiteq}),
\begin{align}\label{eq:hmffull}
\ham_{MF}(\bd{q})=&\sum_{\bd{k}} \bd{C}_{\bd{k},\bd{q}}^{\dagger} H_{BdG,\bd{k}}(\bd{q}) \bd{C}_{\bd{k},\bd{q}} + N\sum_{\alpha=1}^s\frac{|\Delta_{\bd{q},\alpha}|^2}{U_{\alpha}} \nonumber \\
&+\sum_\bd{k}\tr\{h_{-\bd{k}+\bd{q}}^\downarrow-\mu_\bd{q}\},
\end{align}
where $\bd{C}_{\bd{k},\bd{q}}=(c_{\bd{k}+\bd{q},\alpha\uparrow},c_{-\bd{k}+\bd{q},\alpha\downarrow}^\dagger)^T$ denotes the Nambu spinor, with $\bd{q}$ half of the CMM, $N$ denotes the number of unit cells and
\be\label{eq:hbdgqapp}
H_{BdG,\bd{k}}(\bd{q})=\begin{pmatrix}
	h^\uparrow_{\bd{k}+\bd{q}}-\mu_\bd{q}&\hat{\Delta}_\bd{q}\\
	\hat{\Delta}^\dagger_\bd{q}&-(h^{\downarrow T}_{-\bd{k}+\bd{q}}-\mu_\bd{q})
\end{pmatrix}
\ee
is the Bogoliubov-de Gennes (BdG) matrix. TRS imposes that $h^{\downarrow T}_{-\bd{k}}=h^\uparrow_\bd{k}$. Chemical potential $\mu_\bd{q}$ and pairing matrix $\hat{\Delta}_\bd{q}$ are attained self-consistently. Pairing matrix $\hat{\Delta}_\bd{q}=\text{diag}\{\Delta_{\bd{q},1},...,\Delta_{\bd{q},s}\}$ in general has its entries complex and $\bd{q}$-dependent.

To compute the SW, we adopt the grand potential method \cite{peotta2015superfluidity}. SW is the second total derivative of the superconducting free energy $F(\bd{q})$ with respect to $\bd{q}$,
\be
\label{eq:fullSWfree}
D_{s,\mu\nu} = \frac{1}{A} \frac{\derd^2 F}{\derd q_\mu \derd q_\nu}\bigg|_{\bd{q}=0}.
\ee
For simplicity, we set the area $A=1$. The free energy $F(\bd{q})=\Omega(\bd{q},\mu_\bd{q},\hat{\Delta}_\bd{q},\hat{\Delta}_\bd{q}^*)+\mu_\bd{q} N_e$, where the grand potential
\begin{align}
\label{eq:GP}
\Omega(\bd{q},\mu_\bd{q},\hat{\Delta}_\bd{q},\hat{\Delta}_\bd{q}^*)=-\frac{1}{\beta}\ln\tr\{e^{-\beta \ham_{MF}(\bd{q})}\}
\end{align}
is a function of $\bd{q},\mu_\bd{q},\hat{\Delta}_\bd{q},\hat{\Delta}_\bd{q}^*$ explicitly. $N_e$ is the fixed average number of electrons. In the presence of TRS, $D_{s,\mu\nu} $ can be related to partial derivatives of the grand potential~\cite{peotta2015superfluidity,chan2022pairing,huhtinen2022revisiting}:
\begin{align}
\label{eq:fullSW}
D_{s,\mu\nu}=\bigg[\frac{\partial^2 \Omega}{\partial q_\mu\partial q_\nu}-  \frac{\derd \Delta_{\bd{q},\alpha}^I}{\derd q_\mu} \frac{\partial^2\Omega}{\partial \Delta_{\bd{q},\alpha}^I\partial \Delta_{\bd{q},\beta}^I}\frac{\derd \Delta_{\bd{q},\beta}^I}{\derd q_\nu}\bigg]\bigg|_{\bd{q}=0} ,
\end{align}
where $\Delta_{\bd{q},\alpha}^I$ is the imaginary part of $\Delta_{\bd{q},\alpha}$, and we have employed the Einstein summation rule for $\alpha,\beta=1,...,s$. $U(1)$ symmetry allows one to fix $\Delta_{\bd{q},1}$ to be real and positve for all $\bd{q}$, so the sum over $\alpha,\beta$ above can be reduced as from $2$ to $s$ \cite{huhtinen2022revisiting}. However, we assume an arbitrary global phase to keep the discussion general.

The first term of Eq. \eqref{eq:fullSW} ($D_s^{(1)}$) contains two contributions to the SW of multiorbital superconductors---the conventional contribution, which is inversely proportional to the effective mass, and the quantum geometric contribution. This quantum geometric SW is proportional to an integral of the quantum metric~\cite{peotta2015superfluidity}, which is lattice geometry-dependent \cite{simon2020}. However, as expected from general considerations, this geometry-dependence is canceled by the second term of Eq.~\eqref{eq:fullSW} (the lattice geometric term, $D_s^{(2)}$), which encodes the order parameter's variation with orbital positions \cite{chan2022pairing,huhtinen2022revisiting,herzog2022manybody}.

\begin{figure*}[t!]
\includegraphics[width=0.98\textwidth]{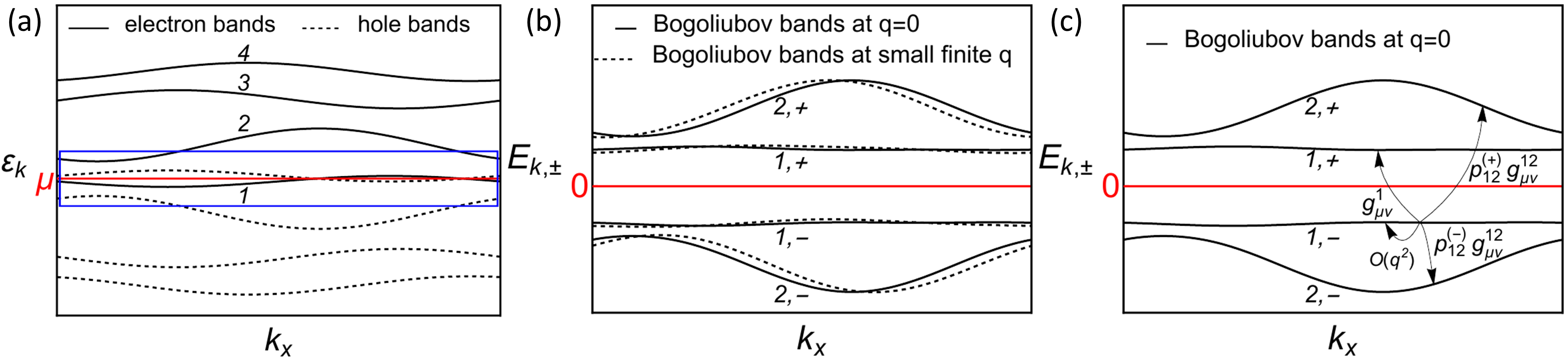}
\caption{Interpretation of the band projection, composite, and geometric terms of Eq. \eqref{eq:d1t} from Bogoliubov band transitions (see App. \ref{app:perturb} for details). (a) Two electronic bands (1,2) energetically close to the interaction scale are projected from a four-orbital model, defining the composite. The blue rectangle indicates the interaction window of $2\Delta$ around the Fermi level. (b)-(c), zoom-in picture of the composite. (b) Electronic bands mix with hole bands to form four Bogoliubov bands ($1,\pm$ and $2,\pm$). At $\bd{q}=0$, these bands are particle-hole symmetric (PHS); at $\bd{q}\neq 0$, the dispersion is deformed and PHS is broken. SW measures the second derivative of energy with respect to $\bd{q}$. (c) For small $\bd{q}$, the deformation can be treated with perturbation, which enables transitions between Bogoliubov bands. Here, only the transitions from band $1,-$ are indicated by arrows. The $O(q^2)$ transition amplitude to band $1,-$ itself, combined with the $O(q)$ amplitude to band $1,+$ gives the intra-band geometric SW $\propto g^{1}_{\mu\nu}$; the $O(q)$ transition amplitudes to band $2,\pm$ give the inter-band terms $\propto p^{(\pm)}_{12}g^{12}_{\mu\nu}$, respectively.}
\label{fig:transition}
\end{figure*}

We consider the scenario that $n \les s$ bands are energetically within or close to the interaction window \cite{footnote1} around the Fermi level (the case of $n=2$ is shown in Fig. \ref{fig:transition} (a)). We refer to the group of $n$ flat or dispersive bands, whether connected or disconnected from each other, as a ``composite", and denote it by $\mathcal{C}$. We perform a projection of the full Hamiltonian Eq. \eqref{eq:hmffull} to the composite, a generalization of the single-band projection in an earlier paper \cite{jiang2023}. The projected mean-field Hamiltonian at finite $\bd{q}$, involving the kinetic energy of the $n$ bands and pairing interactions between them, reads
\begin{widetext}
\begin{align}\label{eq:hcq}
\ham_\mathcal{C}(\bd{q})=&\sum_{\bd{k}}\sum_{l\in\mathcal{C}}[\xi^\uparrow_{l,\bd{k}+\bd{q}}c^\dagger_{l,\bd{k}+\bd{q}\uparrow}c_{l,\bd{k}+\bd{q}\uparrow}-\xi^\downarrow_{l,-\bd{k}+\bd{q}}c_{l,-\bd{k}+\bd{q}\downarrow}c^\dagger_{l,-\bd{k}+\bd{q}\downarrow}]+\sum_\bd{k}\sum_{l\in\mathcal{C}}\xi^\downarrow_{l,-\bd{k}+\bd{q}} \nonumber\\
&+\sum_\bd{k}\sum_{l,l'\in\mathcal{C}}[\Delta_{ll',\bd{k}}(\bd{q})c^\dagger_{l,\bd{k}+\bd{q}\uparrow}c^\dagger_{l',-\bd{k}+\bd{q}\downarrow}+h.c.]+N\sum_{\alpha=1}^s\frac{|\Delta_{\bd{q},\alpha}|^2}{U_{\alpha}},
\end{align}
\end{widetext}
where $\xi^\uparrow_{l,\bd{k}+\bd{q}}=\varepsilon^\uparrow_{l,\bd{k}+\bd{q}}-\mu_\bd{q}$,  $\xi^\downarrow_{l,-\bd{k}+\bd{q}}=\varepsilon^\downarrow_{l,-\bd{k}+\bd{q}}-\mu_\bd{q}$ are the band energies measured from the fluctuated chemical potential. By TRS we have defined $\xi_{l\bd{k}}\equiv\xi^\uparrow_{l\bd{k}}=\xi^\downarrow_{l,-\bd{k}}$. Projection to the composite naturally defines the inter-band gap functions
\begin{align}
\label{eq:interbandgap}
\Delta_{ll',\bd{k}}(\bd{q})\equiv\la u_{l,\bd{k}+\bd{q}}|\hat{\Delta}_\bd{q}|u_{l',\bd{k}-\bd{q}}\ra,
\end{align}
where $l,l'$ are band indices restricted in the composite and $u_{l\bd{k}}$ is the periodic part of Bloch function of the $l^{th}$ spin-$\uparrow$ band. For the case $l=l'$, $\Delta_{ll',\bd{k}}(\bd{q})$ is just the intra-band gap function $\Delta_{l,\bd{k}}(\bd{q})$ \cite{jiang2023}. Under the UPC, the inter-band gap functions ($l\neq l'$) vanish at $\bd{q}=0$ by the orthogonality of Bloch functions but are nonvanishing at $\bd{q}\neq 0$. After projection to the composite, the projected order parameter has equal pairing on the orbitals that make up the composite. In addition, our projection formalism also applies to a general case of intra-orbital pairing when the pairing matrix takes the form $\hat{\Delta}=\text{diag}\{\Delta_1,...,\Delta_1,\Delta_2,...,\Delta_2,...\}$ (i.e., uniform among subsets of orbitals). By definition of the gap functions, Eq. \eqref{eq:interbandgap}, the pairing between band $l$ and $l'$ in the composite can be viewed as uniform if they are spanned by orbitals in a single subset. Therefore, we only require equal pairing on the orbitals of the composite bands.

We developed a stationary state perturbation method to diagonalize the quadratic Hamiltonian of composite bands, Eq. \eqref{eq:hcq}. The perturbation method has been commonly used to simplify Kubo formula results \cite{tinkham2004introduction} but is mainly restricted to the single-band case without considering band geometry. Here, it not only reproduces the first term of Eq. \eqref{eq:fullSW}, $D_s^{(1)}$, which agrees with the general results in Ref. \cite{peotta2015superfluidity,liang2017band}, but gives each contribution the meaning of transitions between Bogoliubov bands.

The first term of Eq. \eqref{eq:fullSW} is (see App. \ref{app:perturb} for details)
\begin{widetext}
\begin{align}\label{eq:d1t}
D^{(1)}_{s,\mu\nu}=&\sum_\bd{k}\sum_{l\in\mathcal{C}}\frac{\Delta^2}{E_{l\bd{k}}^3}\bigg(\tanh\frac{\beta E_{l\bd{k}}}{2}-\frac{\beta E_{l\bd{k}}}{2}\text{sech}^2\frac{\beta E_{l\bd{k}}}{2}\bigg)\partial_\mu \xi_{l\bd{k}}\partial_\nu\xi_{l\bd{k}} \nonumber\\
&+\sum_\bd{k}\sum_{l\in\mathcal{C}}\tanh\frac{\beta E_{l\bd{k}}}{2}\bigg\{\frac{4\Delta^2}{E_{l\bd{k}}}g^l_{\mu\nu}(\bd{k})+\sum_{l'\in\mathcal{C},l'\neq l}8\Delta^2\bigg[\frac{p^{(+)}_{ll'}(\bd{k})}{E_{l\bd{k}}+E_{l'\bd{k}}}+\frac{p^{(-)}_{ll'}(\bd{k})}{E_{l\bd{k}}-E_{l'\bd{k}}}\bigg]g^{ll'}_{\mu\nu}(\bd{k})\bigg\},
\end{align}
\end{widetext}
where $\Delta$ is the uniform pairing order parameter at finite temperature $T=1/\beta$, and $E_{l\bd{k}}=\sqrt{\xi_{l\bd{k}}^2+\Delta^2}$ is the $\bd{q}=0$ quasiparticle energy. The first line of Eq. \eqref{eq:d1t} corresponds to the conventional contribution to SW, and the two separate pieces in the second line correspond to the intra- and inter-band geometric contributions. The different transition processes represented by these geometric contributions can be interpreted from stationary state perturbations and are illustrated in Fig. \ref{fig:transition} (see App. \ref{app:perturb} for details). We have introduced functions
\begin{align}
\label{eq:pfactor}
p^{(\pm)}_{ll'}(\bd{k})\equiv\frac{1}{2}\bigg(1\pm\frac{\xi_{l\bd{k}}\xi_{l'\bd{k}}+\Delta^2}{E_{l\bd{k}}E_{l'\bd{k}}}\bigg),\,\,\,l\neq l',
\end{align}
which can be termed as the $\bd{q}=0$ \tit{inter-band coherence factors} \cite{kopnin2011surface}, in contrast to the intra-band coherence factors ($l=l'$) in conventional BCS theory \cite{schrieffer1999theory,tinkham2004introduction}. $g^l_{\mu\nu}=\re R_{\mu\nu}^{l}$, $g^{ll'}_{\mu\nu}=\re R_{\mu\nu}^{ll'}$ are the real part of the intra- and inter-band component of the quantum geometric tensor $R_{\mu\nu}^{l},R_{\mu\nu}^{ll'}$, respectively, with
\begin{align}\label{eq:glldef1}
&R_{\mu\nu}^{l}(\bd{k})=\la \partial_{\mu} u_{l\bd{k}} |(1- |u_{l\bd{k}} \ra \la u_{l\bd{k}} |)|\partial_{\nu} u_{l\bd{k}} \ra, \nonumber\\
&R^{ll'}_{\mu\nu}(\bd{k})=-\la\partial_\mu u_{l\bd{k}}|u_{l'\bd{k}}\ra\la u_{l'\bd{k}}|\partial_\nu u_{l\bd{k}}\ra,\,\,\,l\neq l'.
\end{align}
They together constitute the quantum geometric tensor of the composite bands,
\begin{align}\label{eq:rcomp}
R^\mathcal{C}_{\mu\nu}(\bd{k})=\sum_{l\in\mathcal{C}}R^l_{\mu\nu}(\bd{k})+\sum_{l,l'\in\mathcal{C},l\neq l'}R^{ll'}_{\mu\nu}(\bd{k}).
\end{align}
where $R^\mathcal{C}_{\mu\nu}(\bd{k})=\tr\{\hat{P}_\mathcal{C}(\bd{k})\partial_\mu\hat{P}_\mathcal{C}(\bd{k})\partial_\nu \hat{P}_\mathcal{C}(\bd{k})\}$ and $\hat{P}_\mathcal{C}(\bd{k})=\sum_{l\in\mathcal{C}}|u_{l\bd{k}}\ra\la u_{l\bd{k}}|$ is the projection operator.

Tensor $g^{ll'}_{\mu\nu}(\bd{k})$ measures small distances between Bloch functions of different bands in the momentum space,
\begin{align}
\label{eq:glldef2}
|\la u_{l\bd{k}}|u_{l',\bd{k}+\bd{q}}\ra|^2\approx-q_\mu q_\nu g^{ll'}_{\mu\nu}(\bd{k}).
\end{align}
With our convention Eq. \eqref{eq:glldef1}, $g^{ll'}_{\mu\nu}(\bd{k})$ is \tit{negative} semidefinite, indicating that the inter-band geometric contributions in Eq. \eqref{eq:d1t} are negative and reduce the total SW. However, in App. \ref{app:pd} we show that the total SW, which is the sum of the conventional, intra- and inter-band geometric terms, is always positive semidefinite. This implies that the $\bd{q}=0$ BCS state locally minimizes the free energy at any temperature whenever the UPC is valid for the composite bands.

In addition to the first term $D^{(1)}_s$, the perturbation method also gives an analytical expression of the Hessian matrix $\partial^2\Omega/\partial \Delta_{\bd{q},\alpha}^I\partial \Delta_{\bd{q},\beta}^I|_{\bd{q}=0}$, thereby put the lattice geometric term $D_s^{(2)}$ into a computationally convenient form. We address this in Sec. \ref{sec:mqm} and \ref{sec:newformula}.

\section{Topological Lower Bound for an Isolated Composite of Flat Bands}
\label{sec:topobound}
The inter-band geometric terms in Eq. \eqref{eq:d1t} are negative semidefinite and reduce the SW. In this section, we use this reduction effect along with the relative topology of the bands in the composite to establish a topological lower bound for the SW of composite bands. For a single isolated flat band, the absolute value of the Chern number provides a weak lower bound for the geometric SW \cite{peotta2015superfluidity}. These bounds stem from the relation
\begin{align}\label{eq:gandb}
\tr g^m({\bd{k}}) \ges |B^m({\bd{k}})|,
\end{align}
for an isolated band $m$, with $B^m$ the Berry curvature of the band, and the fact that the gauge-invariant part of the Wannier localization functional (WLF) is related to the trace of quantum metric. Additional lower bounds due to Wilson loop winding numbers or real space invariants in other topological classifications have also been established \cite{tovmasyan2016effective,xie2020topology,herzog2022superfluid}. One might ask, for composite bands whether there is a lower bound for the SW, and what are the possible generalizations of Eq. \eqref{eq:gandb}?

To answer these questions, we restrict our discussion to the zero temperature limit in the following. At $T=0$, the problem is simplified and Eq. \eqref{eq:d1t} reads
\begin{align}\label{eq:dt0}
D^{(1)}_{s,\mu\nu}=&\sum_{\bd{k}}\sum_{l\in\mathcal{C}} \bigg[\frac{\Delta^2}{E_{l\bd{k}}^3} \partial_\mu \xi_{l\bd{k}}\partial_\nu\xi_{l\bd{k}}+\frac{4\Delta^2}{E_{l\bd{k}}}g_{\mu\nu}^{l}(\bd{k})\bigg] \nonumber\\
&+\sum_\bd{k}\sum_{l,l'\in\mathcal{C},l'>l}\frac{16\Delta^2 }{E_{l\bd{k}}+E_{l'\bd{k}}}p^{(+)}_{ll'}(\bd{k})g^{ll'}_{\mu\nu}(\bd{k}).
\end{align}
Here, coherence factors $p^{(-)}_{ll'}$ are canceled because they describe Bogoliubov transitions between two quasiparticle bands or two quasihole bands [Fig. \ref{fig:transition}(c)], which are suppressed at $T=0$, a ``Pauli blocking" effect. Following a similar procedure described in App. \ref{app:pd}, we break the intra-band quantum metric $g^l$ into inter-band ones, then the geometric part of Eq. \eqref{eq:dt0} splits into two terms,
\begin{align}\label{eq:separation}
D_{s,\mu\nu}^{(1)\text{geo}}=&-\sum_\bd{k}\sum_{l\in\mathcal{C},l'\notin\mathcal{C}}\frac{4\Delta^2}{E_{l\bd{k}}}g^{ll'}_{\mu\nu}(\bd{k})\\
&-\sum_\bd{k}\sum_{l,l'\in\mathcal{C},l'>l}\frac{4\Delta^2(\xi_{l\bd{k}}-\xi_{l'\bd{k}})^2}{E_{l\bd{k}}E_{l'\bd{k}}(E_{l\bd{k}}+E_{l'\bd{k}})}g^{ll'}_{\mu\nu}(\bd{k}). \nonumber
\end{align}
In this expression, the first term contains the inter-band quantum metric between bands inside and outside the composite, while the second term contains these between bands inside the composite only.

To establish a topological lower bound, we consider a \tit{special class} of composite bands, where each band in the composite is within the interaction window, and these bands are energetically close to each other---we call this an ``isolated composite of flat bands"( ICFB). Hereafter, our definition of flat bands encompasses weakly dispersive bands. This requires that the bandwidth of each band, $\delta\xi_{l}=\max\{|\xi_{l\bd{k}}-\xi_{l\bd{k}'}|\}$, and the band gaps between them, $\delta\xi_{ll'}=\max\{|\xi_{l\bd{k}}-\xi_{l'\bd{k}'}|\}$ ($l,l'\in\mathcal{C}$) are all much smaller than the uniform order parameter $\Delta$. This notion of ICFB gives the separation in Eq. \eqref{eq:separation} a well-defined physical meaning.

In Eq. \eqref{eq:separation}, the first term depends on the overall topology of the composite since it contains the quantum metric between bands inside and outside the composite, whereas the second term depends on the relative topology between bands inside the composite. For an ICFB, as long as the first term is nonzero, it is of order $\Delta$ and dominates over the second term, which is of order $\delta\xi^2/\Delta$.

Furthermore, we assume that the ICFB contains an incomplete set of $n$ disconnected bands ($n<s$), each with Chern number $C_l$ ($1\les l\les n$). When a few bands are connected, the Chern numbers will be generalized to the integrals of nonabelian Berry curvature over the Brillouin zone \cite{bohm2003geometric}. The total Chern number of an ICFB, $C=\sum_{l=1}^n C_l$, is always an integer.

With these clarifications, it is important to note that the inequality Eq. \eqref{eq:gandb} holds for composite bands also \cite{peotta2015superfluidity,marzari1997,xie2020topology,ozawa2021relations,mera2022nontrivial}, which reads
\begin{align}\label{eq:gandb2}
\tr g^{\mathcal{C}}({\bd{k}}) \ges |B^{\mathcal{C}}({\bd{k}})|,
\end{align}
where $g^{\mathcal{C}}$ and $B^{\mathcal{C}}$ are the quantum metric and Berry curvature of the composite. Following the definition Eq. \eqref{eq:rcomp}, we write the quantum metric of composite bands as
\begin{align}\label{eq:qgexpandsion}
g^{\mathcal{C}}_{\mu\nu}=\sum_{l\in\mathcal{C}} g^{l}_{\mu\nu}+\sum_{l,l'\in\mathcal{C},l'\neq l} g^{ll'}_{\mu\nu}=-\sum_{l\in\mathcal{C},l'\notin\mathcal{C}}g^{ll'}_{\mu\nu}.
\end{align}
This is proportional to the leading term of Eq. \eqref{eq:separation}, since $\Delta^2/E_{l\bd{k}}$ is approximately a constant for an ICFB. As a result, the total SW is lowered bounded by $|C|$ (besides a proportional constant), with $C$ the total Chern number of the composite. Exceptional cases may exist, e.g., when the composite is complete ($n=s$), such that $C=0$ and the first term of Eq. \eqref{eq:separation} vanishes. In these cases, the SW will rely on the second term of Eq. \eqref{eq:separation} and the conventional term, which depends on the dispersion details; then, a simple topological lower bound cannot be obtained. We will see examples below when the composite has $C=0$ but still provides a nonzero topological bound.

The result $D_s\geqslant |C|$ can be compared to the single isolated band case. If we add the lower bound $|C_l|$ of each band, it would give $D_s\geqslant \sum_{l=1}^n|C_l|$ for a composite. Therefore, we deduce that by the reduction effect of the inter-band geometric SW, the lower bound shrinks from the sum of absolute values to the absolute value of sum, $|C|=|\sum_{l=1}^nC_l|$. This poses a contrast between composites with $\{C_l\}$ of uniform signs and different signs, e.g., $+,+,...$ and $+,-,...$, which can be further explained by analyzing the WLF \cite{marzari1997,peotta2015superfluidity}.

The absolute value of Chern number of the composite bands, $|C|$, is a lower bound for the WLF of Wannier orbitals of the composite, providing a measure of Wannier obstruction. If the composite is Wannier representable, then the bands inside and outside the composite can be constructed from disjoint sets of Wannier orbitals. In the language of ``topological quantum chemistry" \cite{bradlyn2017,cano2018building,zak1980symmetry,zak1982band}, such a composite is said to form a BR. In this case, the inter-band terms of Eq. \eqref{eq:dt0} cancel the intra-band terms altogether, giving a zero lower bound for the leading term of Eq. \eqref{eq:separation}. On the contrary, in the presence of Wannier obstruction, the composite has to form a BR with some bands outside the composite. Then the inter-band terms of Eq. \eqref{eq:dt0} partially cancel the intra-band terms, leaving a finite lower bound for the leading term of Eq. \eqref{eq:separation}.
\begin{figure}[t!]
\includegraphics[width=0.49\textwidth]{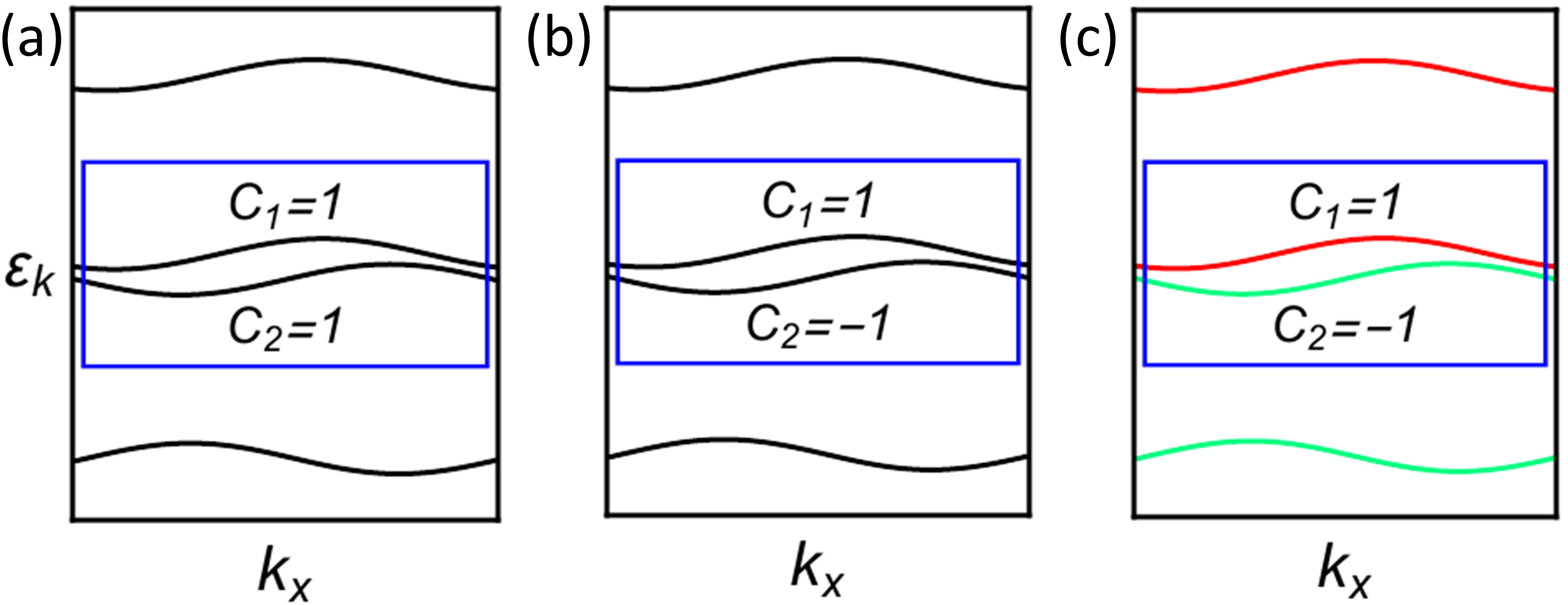}
\caption{An isolated composite of two weakly dispersive bands near the Fermi level (only spin-up bands are shown), with the blue rectangle indicating the interaction window of $2\Delta$. (a) The two bands in the composite have Chern number $C_1=C_2=1$. The SW is lower bounded by $|C|=2$. (b) The two bands have $C_1=1$ and $C_2=-1$, and the SW is lower bound by $|C|=0$. (c), same as (b), but the two bands belong to different EBRs (shown in red and green), then the lower bound can be improved to $|C_1|+|C_2|=2$.}
\label{fig:topobound}
\end{figure}

However, the lower bound by $|C|$ in some cases is too weak, and a stronger lower bound can be derived. Suppose the $s$ bands formed by the $s$ atomic orbitals are divided into a few EBRs \cite{bradlyn2017,zak1980symmetry}, which are constructed from disjoint sets of Wannier basis. For each EBR, it might be that only part of the bands belong to the composite. By making an induction from BR to EBR, one can immediately deduce that the inter-band geometric terms of Eq. \eqref{eq:dt0} can only reduce the SW substantially within an EBR, and not between two EBRs \cite{footnote2}. As a result, the topological lower bound can be improved, with $|C|$ replaced by
\begin{align}\label{eq:improvebound}
\bigg|\sum_{l_1=1}^{n_1}C^{(1)}_{l_1}\bigg|+...+\bigg|\sum_{l_J=1}^{n_J}C^{(J)}_{l_J}\bigg|=\sum_{j=1}^J\bigg|\sum_{l_j=1}^{n_j}C^{(j)}_{l_j}\bigg|,
\end{align}
where $J$ is the number of EBRs formed by the $s$ atomic orbitals, $n_j$ is the number of bands in the composite that is from the $j^{th}$ EBR (with the sum $\sum_{j=1}^J n_j=n$), and $C^{(j)}_{l_j}$ is the Chern number of $l_j\,^{th}$ band of the $j^{th}$ EBR. In the extreme case when each band of the composite belongs to a different EBR, there is no substantial reduction of SW by the inter-band terms, so the improved lower bound is $\sum_{l=1}^n|C_l|$. This simple idea of reducing topological lower bound within individual EBRs is illustrated in Fig. \ref{fig:topobound} using the example of a two-band composite. Similar arguments can be generalized to systems with nonzero Chern numbers to attain stronger bounds based on real space invariants \cite{herzog2022superfluid}.

%
\section{Superfluid Weight and Lattice Geometry}
\label{sec:mqm}
We now study the relationship between SW and lattice geometry using the perturbation method. Here ``lattice geometry" refers to the position of atomic orbitals in a unit cell of non-Bravais lattices. It was shown in Ref. \cite{huhtinen2022revisiting} that when the atomic orbitals occupy some optimal positions, the second term of SW $D_s^{(2)}$ vanishes and the geometric part of $D_s^{(1)}$ becomes the minimal quantum metric (MQM). Typically, the optimal positions are high-symmetry points in the real space lattice.

However, if the symmetry of the tight-binding Hamiltonian is lower than that of the underlying Bravais lattice, the optimal positions can shift along high-symmetry lines \cite{huhtinen2022revisiting}. Since Bravais lattice symmetries no longer protect them, the optimal positions can be sensitive to temperature, interaction, and the tight-binding graph $t_{ij,\alpha\beta}^\sigma$. This observation suggests that instead of seeking the optimal positions of MQM, it would be more straightforward to work with some general positions and find an explicit expression of the second term $D_s^{(2)}$. Notice in Eq. \eqref{eq:fullSW}, one has to either numerically compute the derivative of order parameters $\derd_\mu\Delta^I_\alpha$ (a short-hand notation of $\derd\Delta^I_{\bd{q},\alpha}/\derd q_\mu|_{\bd{q}=0}$) or vary the first term $D_s^{(1)}$ with respect to orbital positions to achieve MQM, both of which may require a large number of calculations.

In this section, we derive an explicit expression of the Hessian matrix $\partial^2\Omega/\partial \Delta_\alpha^I\partial \Delta_\beta^I$ (similarly, stands for $\partial^2\Omega/\partial \Delta_{\bd{q},\alpha}^I\partial \Delta_{\bd{q},\beta}^I|_{\bd{q}=0}$) in Eq. \eqref{eq:fullSW} for general composite bands. It also contains an analysis of its rank and semidefiniteness, which allows us to write $D_s^{(2)}$ into a convenient form in Sec. \ref{sec:newformula}.

Before we proceed, it is essential to clarify and distinguish between geometry-independent and dependent quantities \cite{simon2020}. Electronic band energies and topological properties, such as Chern numbers, are determined by the hopping integrals $t^{\sigma}_{ij,\alpha\beta}$, therefore, are geometry-independent. Geometry-dependent quantities, like the quantum metric and Berry curvature, instead depend on the orbital positions. Following Ref. \cite{simon2020}, we adopt the convention that the atomic orbitals are point-like, therefore the periodic part of the Bloch function, $u_{l\bd{k}}$ depends on the orbital positions through a component-wise phase, $u_{l\bd{k}\alpha}=e^{-i\bd{k}\cdot\bd{x}_\alpha}u^0_{l\bd{k}\alpha}$. Here $u^0_{l\bd{k}\alpha}$ are the Bloch components of gauge choice at some arbitrary reference positions in the unit cell, and $\bd{x}_\alpha$ are the orbital positions measured from these reference points. This explains the origin of geometric dependence of $D_s^{(1)}$, Eq. \eqref{eq:d1t}.

According to the spirit of Ref.~\cite{huhtinen2022revisiting}, the MQM can be defined in two ways, which are precisely given below:\\
\tit{Definition 1:} MQM corresponds to the orbital positions that make the second term of Eq. \eqref{eq:fullSW}, $D_s^{(2)}$ vanish.\\
\tit{Definition 2:} MQM corresponds to the orbital positions that extremize (in fact, minimize) some geometric functional of the quantum metric tensor, which in general takes the form
\begin{align}
\label{eq:functional2}
I=\sum_\bd{k}\big[\sum_{l\in\mathcal{C}}f_l(\bd{k})\tr g^l(\bd{k})+\sum_{l,l'\in\mathcal{C},l\neq l'}f_{ll'}(\bd{k})\tr g^{ll'}(\bd{k})\big].
\end{align}
$I$ is a function of orbital positions $\bd{x}_1,...,\bd{x}_s$ because the quantum metric $g^l_{\mu\nu}(\bd{k}),g^{ll'}_{\mu\nu}(\bd{k})$ depend on them explicitly. $f_l(\bd{k})$ and $f_{ll'}(\bd{k})$ are a set of intra- and inter-band geometry-independent functions. For finite-temperature SW calculations, they are
\begin{align}\label{eq:weight1}
f_l(\bd{k})=\frac{1}{E_{l\bd{k}}}\tanh\frac{\beta E_{l\bd{k}}}{2},
\end{align}
\begin{align}\label{eq:weight2}
f_{ll'}(\bd{k})=\bigg[\frac{2p^{(+)}_{ll'}(\bd{k})}{E_{l\bd{k}}+E_{l'\bd{k}}}+\frac{2p^{(-)}_{ll'}(\bd{k})}{E_{l\bd{k}}-E_{l'\bd{k}}}\bigg]\tanh\frac{\beta E_{l\bd{k}}}{2},
\end{align}
such that the functional $I$ is proportional to the trace of the geometric part of $D^{(1)}_s$. Notice that the quasiparticle energy and coherence factors depend on the band energy only, therefore are geometry-independent.

The equivalence of the two definitions above relies on an assumption made in Ref. \cite{huhtinen2022revisiting} about the rank and semidefiniteness of matrix $\partial^2\Omega/\partial \Delta_\alpha^I\partial \Delta_\beta^I$ (see App. \ref{app:equivalence}). Below, we prove this assumption by computing the matrix explicitly.

It was argued in Ref. \cite{huhtinen2022revisiting} that the order parameter minimizes the superconducting free energy. Therefore, the matrix $\partial^2\Omega/\partial\Delta^I_\alpha\partial\Delta^I_\beta$ has to be positive semidefinite. However, this statement may not be true for a general pairing matrix $\hat{\Delta}$ as a solution to the gap equation of multiorbital superconductors. As a solution channel, it guarantees that the first derivative of grand potential $\Omega$ with respect to $\hat{\Delta}$ is zero (see App. \ref{app:selfconsistency}), but in general cannot imply information about the second derivative. The second derivative instead is related to the ``stability" of the channel.
\begin{figure}[t!]
\includegraphics[width=0.25\textwidth]{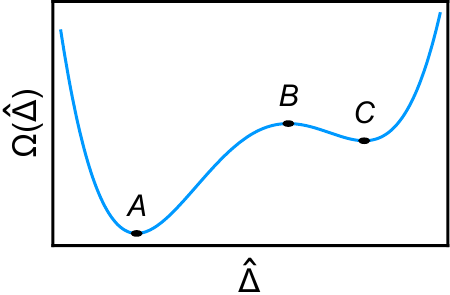}
\caption{Schematic plot of grand potential in the space of components of $\hat{\Delta}$, at $\bd{q}=0$. multiorbital superconductors may have multiple solution channels, $A, B, C$, etc., to some fixed interaction parameters. A solution channel $\hat{\Delta}$ satisfies the gap equation $\partial\Omega/\partial\hat{\Delta}=0$, so extremizes $\Omega$, but the minimization is not guaranteed. A ground state channel (point $A$) ensures the global minimum, but there may also exist ``unstable" solutions (e.g., point $B$). This instability is irrelevant to convergence for numerically solving the gap equation.}
\label{fig:grandplot}
\end{figure}

A multiorbital superconductor can have multiple channels that are solutions to some fixed interaction parameters, an example of which can be seen in the supplementary material of Ref. \cite{jiang2023}. If some channel corresponds to the superconducting ground state, it should be the global minimum of free energy. Therefore, its Hessian matrix needs to be positive semidefinite (Fig. \ref{fig:grandplot}). However, for solutions that are not the ground state, the Hessian matrix may not be positive semidefinite. For instance, in App. \ref{app:npd}, we show for two-orbital models when the order parameter matrix takes the form $\hat{\Delta}=\text{diag}\{\Delta_1,\Delta_2\}$ with $\Delta_1,\Delta_2$ both real, $\partial^2\Omega/\partial\Delta^I_\alpha\partial\Delta^I_\beta$ is positive semidefinite if and only if $\sgn(\Delta_1\Delta_2)=1$. For this reason, obtaining an explicit expression of this matrix for the uniform pairing channel is necessary. To achieve this goal, we study the gap equation for composite bands at finite $\bd{q}$ and extract the Hessian matrix from its derivative with respect to $\bd{q}$.

Let's define the finite-$\bd{q}$ mean-field order parameter as
\begin{align}\label{eq:gapgeneral}
\Delta_{\bd{q},\alpha}=-U_\alpha\la c_{i\alpha\downarrow}c_{i\alpha\uparrow}\ra_\bd{q}e^{-2i\bd{q}\cdot(\bd{R}_i+\bd{x}_\alpha)},
\end{align}
where $\la\ra_\bd{q}$ is the average over the fluctuated pairing state $\Psi_\bd{q}$, and the phase factor $e^{-2i\bd{q}\cdot(\bd{R}_i+\bd{x}_\alpha)}$ gets rid of the superficial $\bd{q}$-dependence from $\Delta_{\bd{q},\alpha}$. Taking the total derivative of Eq. \eqref{eq:gapgeneral} with respect to $q_\mu$ at $\bd{q}=0$, it yields a matrix equation \cite{chan2022pairing,huhtinen2022revisiting}
\begin{align}\label{eq:centraleq}
M_{\alpha\beta}\derd_\mu\Delta_\beta=V_{\alpha,\mu},
\end{align}
where $\derd_\mu\Delta_\alpha=i\derd_\mu\Delta_\alpha^I$ by TRS. After choosing a proportional constant, it is easy to show that (see App. \ref{app:derofgap})
\begin{align}\label{eq:mdef}
M_{\alpha\beta}=\frac{1}{2}\frac{\partial^2\Omega}{\partial \Delta_{\bd{q},\alpha}^I\partial \Delta_{\bd{q},\beta}^I}\bigg|_{\bd{q}=0},
\end{align}
\begin{align}\label{eq:vdef}
V_{\alpha,\mu}=-\frac{\partial^2\Omega}{\partial\Delta_{\bd{q},\alpha}^*\partial q_\mu}\bigg|_{\bd{q}=0},
\end{align}
so the problem of computing matrix $\partial^2\Omega/\partial\Delta^I_\alpha\partial\Delta^I_\beta$ is reduced to determining the gap equation \eqref{eq:gapgeneral}.

For example, the zero-temperature gap equation for a single isolated band $m$ can be expressed as \cite{jiang2023}
\begin{align}\label{eq:gapsingle}
\Delta_{\bd{q},\alpha}=&\frac{U_\alpha}{N}\sum_{\bd{k}}u_{m,\bd{k}-\bd{q},\alpha}^*u_{m,\bd{k}+\bd{q},\alpha} \nonumber\\
&\times\frac{\Delta_{m,\bd{k}}(\bd{q})}{\sqrt{(\xi_{m,\bd{k}+\bd{q}}+\xi_{m,\bd{k}-\bd{q}})^2+4|\Delta_{m,\bd{k}}(\bd{q})|^2}},
\end{align}
from which the derivative can be easily calculated. Here $\Delta_{m,\bd{k}}(\bd{q})=\la u_{m,\bd{k}+\bd{q}}|\hat{\Delta}_\bd{q}|u_{m,\bd{k}-\bd{q}}\ra$ is the projected intra-band gap function. Unfortunately, for general composite bands, a closed form like Eq. \eqref{eq:gapsingle} does not exist since one needs to diagonalize the $2n$-dim mean-field Hamiltonian Eq. \eqref{eq:hcq} to get the pairing state $\Psi_\bd{q}$. However, the perturbation method allows us to expand $\Psi_\bd{q}$ and the gap equation in powers of $\bd{q}$. Thus, Eq. \eqref{eq:centraleq} as its first derivative can be exactly calculated.

To proceed, we transform the gap equation \eqref{eq:gapgeneral} from orbital to band basis and project to the composite, reading
\begin{align}\label{eq:gapcomposite}
\Delta_{\bd{q},\alpha}=-\frac{U_{\alpha}}{N}\sum_{\bd{k}}\sum_{l,l'\in\mathcal{C}} u_{l',\bd{k}-\bd{q},\alpha}^*u_{l,\bd{k}+\bd{q},\alpha}\la c_{l',-\bd{k}+\bd{q}\downarrow}c_{l,\bd{k}+\bd{q}\uparrow}\ra_\bd{q}.
\end{align}
Then we use the standard perturbation method to calculate the pairing amplitude $\la c_{l',-\bd{k}+\bd{q}\downarrow}c_{l,\bd{k}+\bd{q}\uparrow}\ra_\bd{q}$ (see App. \ref{app:amplitude}). Taking total derivative of Eq. \eqref{eq:gapcomposite} with $q_\mu$, we finally obtain Eq. \eqref{eq:centraleq} with
\begin{widetext}
\begin{align}\label{eq:mmatrix}
M_{\alpha\beta}=&\sum_{\bd{k}}\sum_{l\in\mathcal{C}}|u_{l\bd{k}\alpha}|^2\delta_{\alpha\beta}\frac{\tanh(\beta E_{l\bd{k}}/2)}{2E_{l\bd{k}}}-\frac{1}{2}\sum_{\bd{k}}\sum_{l,l'\in\mathcal{C}}u_{l'\bd{k}\alpha}^*u_{l\bd{k}\alpha}u_{l\bd{k}\beta}^*u_{l'\bd{k}\beta} \nonumber \\
&\times\bigg[\frac{p^{(+)}_{ll'}(\bd{k})}{E_{l\bd{k}}+E_{l'\bd{k}}}\bigg(\tanh\frac{\beta E_{l\bd{k}}}{2}+\tanh\frac{\beta E_{l'\bd{k}}}{2}\bigg)+\frac{p^{(-)}_{ll'}(\bd{k})}{E_{l\bd{k}}-E_{l'\bd{k}}}\bigg(\tanh\frac{\beta E_{l\bd{k}}}{2}-\tanh\frac{\beta E_{l'\bd{k}}}{2}\bigg)\bigg]
\end{align}
and
\begin{align}\label{eq:vvector}
V_{\alpha,\mu}=&\sum_{\bd{k}}\sum_{l\in\mathcal{C}}(u_{l\bd{k}\alpha}^*\partial_\mu u_{l\bd{k}\alpha}- \partial_\mu u_{l\bd{k}\alpha}^* u_{l\bd{k}\alpha})\frac{\Delta}{2E_{l\bd{k}}}\tanh\frac{\beta E_{l\bd{k}}}{2}-\sum_{\bd{k}}\sum_{l,l'\in\mathcal{C}}u_{l'\bd{k}\alpha}^*u_{l\bd{k}\alpha}\la u_{l\bd{k}}|\partial_\mu u_{l'\bd{k}}\ra \nonumber\\
&\times \bigg[\frac{\Delta p^{(+)}_{ll'}(\bd{k})}{E_{l\bd{k}}+E_{l'\bd{k}}}\bigg(\tanh\frac{\beta E_{l\bd{k}}}{2}+\tanh\frac{\beta E_{l'\bd{k}}}{2}\bigg)+\frac{\Delta p^{(-)}_{ll'}(\bd{k})}{E_{l\bd{k}}-E_{l'\bd{k}}}\bigg(\tanh\frac{\beta E_{l\bd{k}}}{2}-\tanh\frac{\beta E_{l'\bd{k}}}{2}\bigg)\bigg].
\end{align}
\end{widetext}

Although matrix $M_{\alpha\beta}$ and vector $V_{\alpha,\mu}$ take a complicated form, they have some simple properties we shall discuss now. We assume that the composite $\mathcal{C}$ contains all the $s$ orbitals of the lattice model; therefore, $M_{\alpha\beta}$ is an $s$-dim square matrix. This assumption may not hold for some exceptional cases, e.g., the atomic limit of the single-band projection of two-orbital models (App. \ref{app:2orbital}) and the isolated flat band limit of Lieb lattice, which will be discussed in Sec. \ref{sec:newformula}.

From Eq. \eqref{eq:mmatrix} and \eqref{eq:vvector} above, one first notices that $M_{\alpha\beta}$ is a real symmetric matrix, while $V_{\alpha,\mu}$ is a purely imaginary vector. Moreover, $M_{\alpha\beta}$ contains factors $|u_{l\bd{k}\alpha}|^2$, $u_{l'\bd{k}\alpha}^*u_{l\bd{k}\alpha}$ and quasiparticle energy $E_{l\bd{k}}$, which are all geometry-independent quantities, therefore is geometry-independent, whereas $V_{\alpha,\mu}$ is geometry-dependent. These properties are all governed by Eq. \eqref{eq:centraleq}.

One can also notice that $M_{\alpha\beta}$ of Eq. \eqref{eq:mmatrix} is not invertible, since $\sum_{\beta=1}^s M_{\alpha\beta}=0$ (similarly, $\sum_{\alpha=1}^sV_{\alpha,\mu}=0$), which implies that it has an eigenvector $\bd{v}_0=(1,1,...,1)^T$ with eigenvalue zero. Physically, this is associated with the $U(1)$ symmetry of the order parameter. The existence of such a kernel vector implies that the rank of $M_{\alpha\beta}$ cannot exceed $s-1$. In general, one can prove that under the UPC if the composite bands contain all the $s$ orbitals, then $\text{Rank}(M)=s-1$. We prove this for two special cases in App. \ref{app:rank}.

Nevertheless, one can imagine a process in which some orbitals are removed from the composite, which may arise from topological phase transitions or changes in the interaction scale. Whenever an $r^{th}$ orbital is removed ($1\leqslant r\leqslant s$), $\bd{e}_r=(...,0,1,0,...)^T$ (with the $r^{th}$ component nonzero only) becomes an additional kernel vector of $M_{\alpha\beta}$, lowering $\text{Rank}(M)$ by one. The reason is that the physics will no longer depend on its position as the $r^{th}$ orbital becomes irrelevant to the composite.

Besides, one can show that $M_{\alpha\beta}$ is positive semidefinite, which means it has one zero eigenvalue and $s-1$ positive eigenvalues (see App. \ref{app:rank}). This vital property, as a consequence of UPC, reveals that the uniform pairing channel belongs to these ``stable" solutions to the gap equation like point $A$ and $C$ in Fig. \ref{fig:grandplot}.
\section{Geometry Independence and the Lattice Geometric Term in Terms of Bloch Functions}
\label{sec:newformula}
With the expression of matrix $M_{\alpha\beta}$ and vector $V_{\alpha,\mu}$, it is not difficult to show that each component of the SW tensor is geometry-independent. For this, we write the second term of Eq. \eqref{eq:fullSW} as
\begin{align}\label{eq:d2}
D^{(2)}_{s,\mu\nu}=2M_{\alpha\beta}\derd_\mu\Delta_\alpha \derd_\nu\Delta_\beta,
\end{align}
where $\derd_\mu\Delta_\alpha=i\derd_\mu\Delta_\alpha^I$. Under a geometric transformation which translates orbital positions from $\bd{x}^0_\alpha$ to $\bd{x}_\alpha=\bd{x}^0_\alpha+\delta \bd{x}_\alpha$ \cite{footnote3}, it is changed by
\begin{align}\label{eq:change2}
\delta D^{(2)}_{s,\mu\nu}=&2M_{\alpha\beta}(\delta \derd_\mu\Delta_\alpha \derd_\nu\Delta_\beta^0+\derd_\mu\Delta_\alpha^0 \delta \derd_\nu\Delta_\beta \nonumber\\
&+\delta \derd_\mu\Delta_\alpha \delta \derd_\nu\Delta_\beta),
\end{align}
where $\derd_\mu\Delta_\alpha^0$ is $\derd_\mu\Delta_\alpha$ evaluated at the initial positions and $\delta \derd_\mu\Delta_\alpha^0$ is the change due to such transformation. Eq. \eqref{eq:change2} can be further simplified using Eq. \eqref{eq:xchange} in App. \ref{app:equivalence}. Similarly, the change of the quantum geometric term of SW, $\delta D^{(1)}_s$, can be evaluated by inserting Eq. \eqref{eq:deltag} into Eq. \eqref{eq:d1t}. Then, one can immediately check the geometry independence of the total SW
\begin{align}\label{eq:independence}
\delta D_s=\delta D^{(1)}_s+\delta D^{(2)}_s=0,
\end{align}
therefore, we have finished establishing the relation between SW and lattice geometry.

As mentioned at the beginning of Sec. \ref{sec:mqm}, a shortcoming of Eq. \eqref{eq:fullSW} to compute the SW is that one needs to either solve the gap equation at a few $\bd{q}$ points to get the derivative $\derd_\mu\Delta_\alpha$, or vary the orbital positions in the whole geometric space to find the MQM positions. However, since we have treated the gap equation perturbatively to get Eq. \eqref{eq:mmatrix} and \eqref{eq:vvector}, it means the total SW can be computed using simple integrals of Bloch functions and their derivatives without following these procedures.

In fact, as pointed out in Ref. \cite{chan2022pairing,huhtinen2022revisiting}, once we are convinced that $\text{Rank}(M)=s-1$, it would be convenient to partially project \cite{footnote4} the matrix $M_{\alpha\beta}$, vector $V_{\alpha,\mu}$ and $\derd_\mu\Delta_\alpha$ into an $(s-1)$-dim subspace, in which the matrix $M$ becomes invertible, so one can write
\begin{align}
\widetilde{\derd_\mu\Delta_\alpha}=(\widetilde{M}^{-1})_{\alpha\beta}\widetilde{V}_{\beta,\mu},
\end{align}
where $\widetilde{}$ denotes the projected quantities. This finally yields a simple expression for the lattice geometric term:
\begin{align}\label{eq:ds2final}
D^{(2)}_{s,\mu\nu}=2(\widetilde{M}^{-1})_{\alpha\beta}\widetilde{V}_{\alpha,\mu}\widetilde{V}_{\beta,\nu}.
\end{align}
In this expression, the projected $\widetilde{M}$ and $\widetilde{V}$ are not unique but can be taken as eliminating the first row and column from matrix $M$ and eliminating the first component from vector $V$ in Eq. \eqref{eq:mmatrix} and \eqref{eq:vvector}. This corresponds to restricting to the geometric space spanned by the $s-1$ orbitals.
\begin{figure}[t!]
\includegraphics[width=0.49\textwidth]{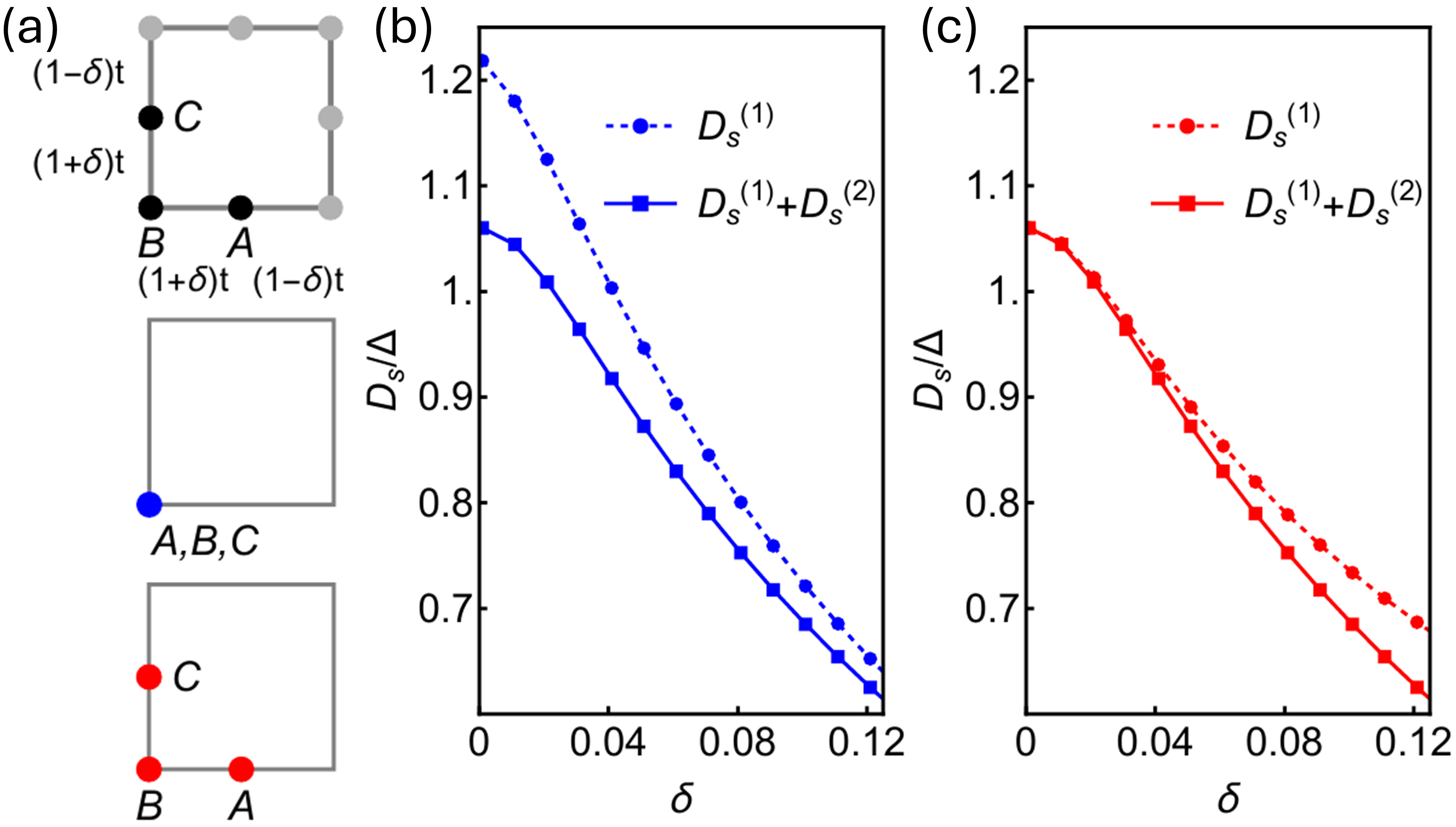}
\caption{SW calculation of the Lieb lattice model. (a) The gray square indicates the unit cell of the Lieb lattice and the three orbitals $A, B, C$. Top: the tight-binding hoppings can be tuned to $(1\pm\delta)t$ to open a gap $2\sqrt{2}t\delta$ between the flat and dispersive bands; Mid: lattice geometry assuming the three orbitals sitting on the same site; Bottom: assuming the orbitals sitting on the regular sites of Lieb lattice. (b)(c) SW computed using Eq. \eqref{eq:d1t} and \eqref{eq:ds2final} at $T=0$, $t/\Delta=10$, $\mu=0$. (b) assumes the geometry of (a) mid while (c) assumes the geometry of (a) bottom.}
\label{fig:lieb}
\end{figure}

To show the utility of Eq. \eqref{eq:ds2final}, we take the Lieb lattice model \cite{julku2016geometric,huhtinen2022revisiting} as an example. A unit cell of the Lieb lattice is shown at the top of Fig. \ref{fig:lieb} (a), where parameters $t$ and $\delta$ completely determine the tight-binding matrix elements. Remember that the hopping graph does not specify the positions of orbital $A, B$, and $C$, which leaves the degree of freedom for writing the Bloch Hamiltonian $h^\sigma_\bd{k}$. Since the hopping integrals are usually defined from a given lattice structure, in this case, one can still imagine the underlying Bravais lattice as a square lattice with $C_{4v}$ symmetry.

When $\delta\neq 0$, the hopping graph breaks $C_4$ symmetry, making the tight-binding model acquire a lower symmetry than the imagined square lattice. Then it was found that the optimal positions of MQM of atom $A$ and $C$ no longer sit at the midpoint of the side but shift towards $B$ atom \cite{huhtinen2022revisiting}. As a contrast, in our calculations using Eq. \eqref{eq:d1t} and \eqref{eq:ds2final}, we can place the orbitals at any general positions in the unit cell. In Fig. \ref{fig:lieb} (b), we imagine the three orbitals sitting on the same site, while Fig. \ref{fig:lieb} (c) assumes they are sitting on the regular sites of the Lieb lattice. Since we make no approximations and take into account all the inter-band and lattice geometric effects, the total SW data of the two geometries (two solid curves in (b) and (c)) completely match, showing the geometry independence of the SW. One also notices that in Fig. \ref{fig:lieb} (b) the $D_s^{(1)}$ curve converges to $D_s^{(1)}+D_s^{(2)}$ in the large $\delta$ limit, whereas in (c) they converge in the $\delta=0$ limit. This exhibits how the actual MQM geometry approaches the two geometries in the two limits.

Finally, we address two possible issues of applying Eq. \eqref{eq:ds2final} for numerical calculations. The first regards the definition of composite for the Lieb lattice. For most parameter ranges, one can take all three bands of the Lieb lattice as the composite, which contains three orbitals, then matrix $\widetilde{M}$ and vector $\widetilde{V}$ are 2-dim. However, in the isolated flat band limit, $\mu\sim 0$ and $\Delta\ll t\delta$, the composite effectively contains just the flat band, which comprises only two orbitals. Then as $\text{Rank}(M)$ is lowered to 1, we need to project $M$ to a 1-dim subspace to make $\widetilde{M}$ invertible. This gives a 1-dim matrix $\widetilde{M}$ and vector $\widetilde{V}$, which means for Eq. \eqref{eq:mmatrix} and \eqref{eq:vvector} we only need to evaluate the integral for orbital $A$ or $C$.

The second is about the differentiability of Bloch functions in calculating vector $V_{\alpha,\mu}$. For systems with non-trivial topology, the Bloch functions may not be differentiable everywhere in the momentum space. Although the derivative of Bloch functions can be locally singular in some cases, one can argue on general grounds that the lattice geometric term $D_s^{(2)}$ is finite. Since the total SW $D_s^{(1)}+D_s^{(2)}$ is geometry-independent and positive finite under the UPC, it implies that $D_s^{(2)}$ is bounded: $-D_s^{(1)}<D_s^{(2)}<0$. Here, we have used the fact that $D_s^{(1)}$, which contains integrals of the quantum metric, is always finite, which can be understood from two generic cases. For gapped insulating states,  the inter-band quantum metric can be written in terms of velocity operators, $\la u_{l\bd{k}}|\partial_\mu u_{l'\bd{k}}\ra=\la u_{l\bd{k}}|\partial_\mu h_\bd{k} |u_{l'\bd{k}}\ra/(\varepsilon_{l'\bd{k}}-\varepsilon_{l\bd{k}})$. Therefore, the derivatives are shifted to the Bloch Hamiltonian, usually a continuous and smooth function of $\bd{k}$. In the case of a band degeneracy, the quantum metric can diverge, giving a $1/k^2$ divergence near the isolated band degeneracy points, leading to a logarithmic divergence of the intra-band contributions to $D_s^{(1)}$. However, this divergence is canceled by the inter-band geometric contributions due to the reduction effect, giving a finite total $D_s^{(1)}$ \cite{iskin2019origin}. In summary, even though the non-differentiability of Bloch functions may cause singularities at some local $\bd{k}$-points, it does not affect the numerical convergence for calculations of both $D_s^{(1)}$ and $D_s^{(2)}$.
\section{Conclusion and Outlook}
\label{sec:discussion}
We developed a perturbation approach to treat multiorbital superconductors' center-of-mass momentum (phase) fluctuation. This approach naturally separates the superfluid weight of composite bands into intra- and inter-band contributions from Bogoluibov band transitions. Based on the reduction by the inter-band geometric terms, a topological lower bound was derived for the superfluid weight of isolated composite flat bands. Applying the theory of band representation, we found that the elementary band representations should be considered the minimum ``block" for deriving a stronger lower bound.

In studying the geometry independence of the superfluid weight, we emphasized the role of the derivative of gap equations in getting the Hessian matrix $\partial^2\Omega/\partial \Delta_\alpha^I\partial \Delta_\beta^I$. The perturbation method facilitated the calculation as we expanded the gap equation in powers of $\bd{q}$. We analyzed the properties of this Hessian matrix in detail and showed how the positive semidefiniteness and rank information stem from the uniform pairing condition. We finally rewrote the superfluid weight formula into a convenient form suitable for numerical calculations of any general tight-binding model.

Since the superfluid weight is the stiffness of the Meissner effect, it naturally encodes the information about fluctuation to a finite-momentum pairing state within a given channel. Still, there is little information about competition between superconducting channels and other correlated states \cite{hofmann2022heuristic,hofmann2023superconductivity}. However, as pointed out, the Hessian matrix of grand potential with respect to order parameters can be nonpositive definite for multiorbital superconductors and contains information about the stability of the channel. Whether this allows one to determine the ground state is left to future studies.\\

\noindent
\tit{Note added}: During the review process, we became aware of Ref. \cite{tam2024geometry}, which discusses the geometry-independence of the superfluid weight for degenerate flat bands using random phase approximation. On topics where there is an overlap, our findings are consistent with theirs.

\acknowledgments{The authors would like to acknowledge Jonah Herzog-Arbeitman for discussions on the EBR and the support of UNR/VPRI startup grant PG19012. G. J. has also been funded by the Jane and Aatos Erkko Foundation and the Keele Foundation as part of the SuperC collaboration.}

\appendix
\section{Interacting Hamiltonian with Center-of-Mass Momentum Fluctuation}
\label{app:finiteq}
By transformation $c_{i\alpha\sigma}=(1/\sqrt{N})\sum_\bd{k}e^{i\bd{k}\cdot(\bd{R}_i+\bd{x}_\alpha)}c_{\bd{k}\alpha\sigma}$ (we chose this gauge for $c_{\bd{k}\alpha\sigma}$ if not otherwise specified), the onsite intra-orbital interaction takes the form
\begin{align}
\ham_{int}&=-\sum_{i,\alpha}U_\alpha c^\dagger_{i\alpha\uparrow}c^\dagger_{i\alpha\downarrow}c_{i\alpha\downarrow}c_{i\alpha\uparrow}\\
&=-\sum_{\bd{k}\bd{k}'\bd{q},\alpha}\frac{U_\alpha}{N}c^\dagger_{\bd{k}+\bd{q},\alpha\uparrow}c^\dagger_{-\bd{k}+\bd{q},\alpha\downarrow}c_{-\bd{k}'+\bd{q},\alpha\downarrow}c_{\bd{k}'+\bd{q},\alpha\uparrow}. \nonumber
\end{align}
The conventional BCS theory makes the ansatz that electron $\bd{k}\uparrow$ pairs with $-\bd{k}\downarrow$, therefore the reduced interaction $\ham_{int}^{(\text{red})}$ only keeps the $\bd{q}=0$ term from above.

We now consider a pairing state of fluctuated CMM $2\bd{q}$ (i.e. phase fluctuation), so the reduced interaction keeps a term of fixed and finite $\bd{q}$, reading
\begin{align}
\ham_{int}^{(\text{red})}(\bd{q})=-\sum_{\bd{k}\bd{k}',\alpha}\frac{U_\alpha}{N}c^\dagger_{\bd{k}+\bd{q},\alpha\uparrow}c^\dagger_{-\bd{k}+\bd{q},\alpha\downarrow}c_{-\bd{k}'+\bd{q},\alpha\downarrow}c_{\bd{k}'+\bd{q},\alpha\uparrow}.
\end{align}
This form of interaction implies that in the fluctuated state electron $\bd{k}+\bd{q},\uparrow$ pairs with $-\bd{k}+\bd{q},\downarrow$. With the order parameter $\Delta_{\bd{q},\alpha}$ defined by Eq. \eqref{eq:gapgeneral}, Fourier transform gives
\begin{align}\label{eq:gapk}
\Delta_{\bd{q},\alpha}=-\frac{U_\alpha}{N}\sum_\bd{k}\la c_{-\bd{k}+\bd{q},\alpha\downarrow}c_{\bd{k}+\bd{q},\alpha\uparrow}\ra_\bd{q}.
\end{align}
To get Eq. \eqref{eq:gapk}, we have used the fact that in the fluctuated state $\Psi_\bd{q}$ the electron $\bd{k}+\bd{q}\uparrow$ pairs with $-\bd{k}+\bd{q}\downarrow$. Using the transformation between orbital and band basis
\begin{align}
&c_{\bd{k}\alpha\uparrow}=\sum_{l}u_{l\bd{k}\alpha}c_{l\bd{k}\uparrow}, \nonumber\\
&c_{-\bd{k},\alpha\downarrow}=\sum_{l}u_{l\bd{k}\alpha}^*c_{l,-\bd{k}\downarrow},
\end{align}
one obtains Eq. \eqref{eq:gapcomposite} (the sum over $l$ has been restricted to the composite due to projection).

With \eqref{eq:gapk}, mean-field decoupling yields
\begin{align}\label{eq:mfint}
\ham_{int}^{(\text{red})}(\bd{q})\simeq\sum_{\bd{k},\alpha}(\Delta_{\bd{q},\alpha}c^\dagger_{\bd{k}+\bd{q},\alpha\uparrow}c^\dagger_{-\bd{k}+\bd{q},\alpha\downarrow}+h.c.)+N\sum_{\alpha=1}^s\frac{|\Delta_{\bd{q},\alpha}|^2}{U_\alpha}.
\end{align}
For the kinetic part, we shift the dummy momentum $\bd{k}$ by $\pm\bd{q}$ for the spin-$\uparrow$, $\downarrow$ sector, respectively,
\begin{widetext}
\begin{align}\label{eq:mfkin}
\ham_{kin}=&\sum_{\bd{k},\alpha\beta}\big[c^\dagger_{\bd{k}\alpha\uparrow}(h^\uparrow_\bd{k})_{\alpha\beta}c_{\bd{k}\beta\uparrow}+c^\dagger_{-\bd{k},\alpha\downarrow}(h^\downarrow_{-\bd{k}})_{\alpha\beta}c_{-\bd{k},\beta\downarrow}\big]-\mu_\bd{q}\sum_{\bd{k},\alpha}(c^\dagger_{\bd{k}\alpha\uparrow}c_{\bd{k}\alpha\uparrow}+c^\dagger_{-\bd{k},\alpha\downarrow}c_{-\bd{k},\alpha\downarrow})\\
=&\sum_{\bd{k},\alpha\beta}\big\{c^\dagger_{\bd{k}+\bd{q},\alpha\uparrow}\big[(h^\uparrow_{\bd{k}+\bd{q}})_{\alpha\beta}-\mu_\bd{q}\delta_{\alpha\beta}\big]c_{\bd{k}+\bd{q},\beta\uparrow} -c_{-\bd{k}+\bd{q},\alpha\downarrow}\big[(h^{\downarrow,T}_{-\bd{k}+\bd{q}})_{\alpha\beta}-\mu_\bd{q}\delta_{\alpha\beta}\big]c^\dagger_{-\bd{k}+\bd{q},\beta\downarrow}\big\}+\sum_\bd{k}\tr\{h^{\downarrow}_{-\bd{k}+\bd{q}}-\mu_\bd{q}\}.  \nonumber
\end{align}
\end{widetext}
In the last line above, $\sum_\bd{k}\tr h^{\downarrow}_{-\bd{k}+\bd{q}}=\sum_{l,\bd{k}}\varepsilon^\downarrow_{l,-\bd{k}+\bd{q}}$ is $\bd{q}$-independent because $\bd{k}$ is summed over the entire Brillouin zone, whereas $\sum_\bd{k}\tr\mu_\bd{q}=\mu_\bd{q}sN$ with $N$ the number of unit cells. Eq. \eqref{eq:mfint} and \eqref{eq:mfkin} together give Eq. \eqref{eq:hmffull}.
\section{Self-consistency Equations}
\label{app:selfconsistency}
For completeness, we derive the gap equation and electron number equation and study their properties under TRS, which is similar to Ref. \cite{peotta2015superfluidity}. With the grand potential Eq. \eqref{eq:GP}, we first derive the gap equation
\begin{align}\label{eq:self1}
\frac{\partial\Omega}{\partial \Delta_{\bd{q},\alpha}^*}=0,\,\,\,\forall\,\bd{q},\alpha.
\end{align}
Using Eq. \eqref{eq:hmffull},
\begin{align}\label{eq:doddelta}
\frac{\partial\Omega}{\partial \Delta_{\bd{q},\alpha}^*}&=\frac{\tr\{e^{-\beta\ham_{MF}(\bd{q})}\sum_\bd{k}c_{-\bd{k}+\bd{q},\alpha\downarrow}c_{\bd{k}+\bd{q},\alpha\uparrow}\}}{\tr\{e^{-\beta\ham_{MF}(\bd{q})}\}}+\frac{N}{U_\alpha}\Delta_{\bd{q},\alpha}\nonumber\\
&=\sum_\bd{k}\la c_{-\bd{k}+\bd{q},\alpha\downarrow}c_{\bd{k}+\bd{q},\alpha\uparrow}\ra_\bd{q}+\frac{N}{U_\alpha}\Delta_{\bd{q},\alpha}.
\end{align}
This cancels by the definition of order parameter \eqref{eq:gapk}; therefore, \eqref{eq:self1} is proved.

Next, we look at the electron number equation:
\begin{align}\label{eq:self2}
\frac{\partial\Omega}{\partial \mu_{\bd{q}}}=-N_e,\,\,\,\forall\,\bd{q},
\end{align}
where the fixed average number of electrons $N_e$ is a constraint for the system; condition $\forall\,\bd{q}$ means when we compare the fluctuated state $\Psi_\bd{q}$ of different $\bd{q}$, $N_e$ is always fixed at the same value. Using Eq. \eqref{eq:hmffull} again one obtains
\begin{align}\label{eq:dodmu}
\frac{\partial\Omega}{\partial \mu_{\bd{q}}}=-\frac{\tr\{e^{-\beta\ham_{MF}(\bd{q})}\sum_{\bd{k}\alpha\sigma}c^\dagger_{\bd{k}\alpha\sigma}c_{\bd{k}\alpha\sigma}\}}{\tr\{e^{-\beta\ham_{MF}(\bd{q})}\}},
\end{align}
therefore Eq. \eqref{eq:self2} is just
\begin{align}\label{eq:numbereq}
N_e=\sum_{\bd{k}\alpha\sigma}\la c^\dagger_{\bd{k}\alpha\sigma}c_{\bd{k}\alpha\sigma}\ra_\bd{q}.
\end{align}

The two self-consistency equations \eqref{eq:gapk}, \eqref{eq:numbereq} determine all the properties of the fluctuated state $\Psi_\bd{q}$, and one can solve them for $\hat{\Delta}_{\bd{q}}$, $\mu_\bd{q}$. In the presence of TRS, one can show that whenever $\{\hat{\Delta}_\bd{q},\mu_\bd{q}\}$ is a solution to the self-consistency equations of $\bd{q}$ state, then $\{\hat{\Delta}_\bd{q}^\dagger,\mu_\bd{q}\}$ is a solution to the equations for $-\bd{q}$ state (the single isolated band case was proved in Ref. \cite{jiang2023}). Then for the case of diagonal $\hat{\Delta}_\bd{q}$ matrix, $\Delta_{-\bd{q},\alpha}=\Delta_{\bd{q},\alpha}^*$ and $\mu_{-\bd{q}}=\mu_\bd{q}$, so there are the following identities:
\begin{align}\label{eq:trs1}
\frac{\derd\mu_\bd{q}}{\derd q_\mu}\bigg|_{\bd{q}=0}=0,
\end{align}
\begin{align}\label{eq:trs2}
\frac{\derd \Delta_{\bd{q},\alpha}}{\derd q_\mu}\bigg|_{\bd{q}=0}&=-\frac{\derd \Delta_{\bd{q},\alpha}^*}{\derd q_\mu}\bigg|_{\bd{q}=0}=i\frac{\derd \Delta_{\bd{q},\alpha}^I}{\derd q_\mu}\bigg|_{\bd{q}=0},
\end{align}
where $ \Delta_{\bd{q},\alpha}^I$ is the imaginary part of $ \Delta_{\bd{q},\alpha}$. The importance of TRS will be discussed further in App. \ref{app:derofgap}.
\section{Superfluid Weight Formula, Derivative of Gap Equation and Time-reversal Symmetry}
\label{app:derofgap}
In this section, we derive the SW formula Eq. \eqref{eq:fullSW}. We treat the free energy $F(\bd{q})$ as function of $\bd{q}$ only, while $\Omega(\bd{q},\mu_\bd{q},\Delta_{\bd{q},\alpha},\Delta_{\bd{q},\alpha}^*)$ is a function of $\bd{q},\mu_\bd{q},\Delta_{\bd{q},\alpha},\Delta_{\bd{q},\alpha}^*$ explicitly.

Using $F(\bd{q})=\Omega(\bd{q},\mu_\bd{q},\Delta_{\bd{q},\alpha},\Delta_{\bd{q},\alpha}^*)+\mu_\bd{q}N_e$, we have the first derivative
\begin{align}
\frac{\derd F}{\derd q_\mu}=&\frac{\partial\Omega}{\partial q_\mu}+\bigg(\frac{\partial\Omega}{\partial \mu_\bd{q}}+N_e\bigg)\frac{\derd \mu_\bd{q}}{\derd q_\mu} \nonumber\\
&+\sum_\alpha\bigg(\frac{\partial\Omega}{\partial \Delta_{\bd{q},\alpha}}\frac{\derd\Delta_{\bd{q},\alpha}}{\derd q_\mu}+c.c.\bigg).
\end{align}
The second term cancels by the electron number equation, and the third term cancels by the gap equation; therefore
\begin{align}
\frac{\derd F}{\derd q_\mu}=\frac{\partial \Omega}{\partial q_\mu},\,\,\,\forall\,\bd{q}.
\end{align}
An important consequence of TRS is that $F(\bd{q})$ is an even function of $\bd{q}$, so
\begin{align}\label{eq:trsf}
\frac{\derd F}{\derd q_\mu}\bigg|_{\bd{q}=0}=\frac{\partial \Omega}{\partial q_\mu}\bigg|_{\bd{q}=0}=0.
\end{align}
Caution is needed if one wants to prove this like Eq. \eqref{eq:doddelta} and \eqref{eq:dodmu},  the derivative should not act on $c_{\bd{k}+\bd{q},\uparrow}$ and $c_{-\bd{k}+\bd{q},\downarrow}$ operators since they are to form the Nambu basis. At the same time, the grand potential depends only on the eigenvalues of the BdG Hamiltonian. One can calculate $\partial\Omega/\partial q_\mu$ directly using an explicit form like Eq. \eqref{eq:grandc}, and then show Eq. \eqref{eq:trsf} holds as long as $\hat{\Delta}_{\bd{q}=0}$ is a Hermitian matrix up to a $U(1)$ phase, which TRS imposes. The case of a single isolated band has been proved in Ref. \cite{jiang2023}. If TRS is broken, SW being the second derivative of free energy at $\bd{q}=0$ will lose its meaning as a stiffness tensor, since the first derivative of free energy is already nonzero.

In the presence of TRS, the second derivative of free energy is
\begin{align}
\frac{\derd^2F}{\derd q_\mu \derd q_\nu}=&\frac{\partial^2 \Omega}{\partial q_\mu\partial q_\nu}+\frac{\partial^2 \Omega}{\partial q_\mu\partial \mu_\bd{q}}\frac{\derd \mu_\bd{q}}{\derd q_\nu} \nonumber\\
&+\sum_\alpha\bigg(\frac{\partial^2 \Omega}{\partial q_\mu\partial \Delta_{\bd{q},\alpha}}\frac{\derd \Delta_{\bd{q},\alpha}}{\derd q_\nu}+c.c.\bigg)
\end{align}
To simplify this, we take the total derivative of gap equation \eqref{eq:self1} with respect to $q_\mu$, yielding
\begin{align}\label{eq:4terms}
&\frac{\partial^2\Omega}{\partial\Delta_{\bd{q},\alpha}^*\partial q_\mu}+\frac{\partial^2\Omega}{\partial\Delta_{\bd{q},\alpha}^*\partial \mu_\bd{q}}\frac{\derd \mu_\bd{q}}{\derd q_\mu} \nonumber\\
&+\sum_\beta\bigg(\frac{\partial^2\Omega}{\partial\Delta_{\bd{q},\alpha}^*\partial \Delta_{\bd{q},\beta}}\frac{\derd \Delta_{\bd{q},\beta}}{\derd q_\mu}+c.c.\bigg)=0.
\end{align}
Changing variables $\Delta_{\bd{q},\alpha},\Delta_{\bd{q},\alpha}^*\rightarrow \Delta_{\bd{q},\alpha}^R,\Delta_{\bd{q},\alpha}^I$ ($\Delta_{\bd{q},\alpha}\equiv \Delta_{\bd{q},\alpha}^R+i\Delta_{\bd{q},\alpha}^I$), evaluating at $\bd{q}=0$ and using TRS properties Eq. \eqref{eq:trs1}, \eqref{eq:trs2}, one can check $\derd^2 F/\derd q_\mu \derd q_\nu|_{\bd{q}=0}$ gives the SW formula Eq. \eqref{eq:fullSW}.

Moreover, Eq. \eqref{eq:4terms} at $\bd{q}=0$ gives
\begin{align}\label{eq:matrixeq}
\frac{1}{2}\frac{\partial^2\Omega}{\partial \Delta_{\bd{q},\alpha}^I\partial \Delta_{\bd{q},\beta}^I}\bigg|_{\bd{q}=0}\derd_\mu\Delta_\beta+\frac{\partial^2\Omega}{\partial\Delta_{\bd{q},\alpha}^*\partial q_\mu}\bigg|_{\bd{q}=0}=0,
\end{align}
which exactly maps to $M_{\alpha\beta}\derd_\mu\Delta_\beta-V_{\alpha,\mu}=0$ and the total derivative of Eq. \eqref{eq:doddelta}. Therefore, we get Eq. \eqref{eq:mdef}, \eqref{eq:vdef} in the main text.
\section{Perturbation Approach to Calculate the Superfluid Weight}
\label{app:perturb}
In this appendix, we provide the steps of using the perturbation method to calculate $D_s^{(1)}$, Eq. \eqref{eq:d1t}.
\subsection{Nondegenerate Perturbation}
Starting from the mean-field Hamiltonian projected to the composite bands, Eq. \eqref{eq:hcq}, the first and third terms can be organized into the $2n$-dim BdG form
\begin{widetext}
\begin{align}\label{eq:bdgmatrix}
H_{\bd{k}}(\bd{q})=\begin{pmatrix}
\xi_{1,\bd{k}+\bd{q}}&\Delta_{1,\bd{k}}(\bd{q})&&\Delta_{12,\bd{k}}(\bd{q})&..&&\Delta_{1n,\bd{k}}(\bd{q})\\
\Delta_{1,\bd{k}}(\bd{q})^*&-\xi_{1,\bd{k}-\bd{q}}&\Delta_{21,\bd{k}}(\bd{q})^*&&..&\Delta_{n1,\bd{k}}(\bd{q})^*&\\
&\Delta_{21,\bd{k}}(\bd{q})&\xi_{2,\bd{k}+\bd{q}}&\Delta_{2,\bd{k}}(\bd{q})&&&\\
\Delta_{12,\bd{k}}(\bd{q})^*&&\Delta_{2,\bd{k}}(\bd{q})^*&-\xi_{2,\bd{k}-\bd{q}}&&&\\
..&..&&&..&&\\
&\Delta_{n1,\bd{k}}(\bd{q})&&&&\xi_{n,\bd{k}+\bd{q}}&\Delta_{n,\bd{k}}(\bd{q})\\
\Delta_{1n,\bd{k}}(\bd{q})^*&&&&&\Delta_{n,\bd{k}}(\bd{q})^*&-\xi_{n,\bd{k}-\bd{q}}
\end{pmatrix},
\end{align}
\end{widetext}
where the empty entries are zeroes. To aid in bookkeeping, in $H_\bd{k}(\bd{q})$ we have assigned the $(2l-1)^{th}$ label to electrons, and the $2l^{th}$ label to holes of the $l^{th}$ band.

Let $\Lambda_{\bd{k}} (\bd{q})$ be some generalized Bogoliubov transformation that exactly diagonalizes $H_\bd{k}(\bd{q})$,
\begin{align}
\Lambda_{\bd{k}} (\bd{q})^\dagger H_\bd{k}(\bd{q})\Lambda_{\bd{k}} (\bd{q})=\hat{E}_\bd{k}(\bd{q}).
\end{align}
Typically $\Lambda_{\bd{k}} (\bd{q})$ can only be determined numerically.
\begin{align}
\hat{E}_\bd{k}(\bd{q})=\text{diag}\{E_{1\bd{k}+}(\bd{q}),E_{1\bd{k}-}(\bd{q}),...\}
\end{align}
is a diagonal matrix consisting eigenvalues of $H_\bd{k}(\bd{q})$ (hat $\,\,\,\hat{}$ is to distinguish from a number). The $2n$ eigenvalues correspond to the $n$ fluctuated Bogoliubov quasiparticle bands ($+$) and $n$ quasihole bands ($-$). At $\bd{q}=0$ the spectrum is particle-hole symmetric (PHS), $E_{l\bd{k}\pm}(0)=\pm\sqrt{\xi_{l\bd{k}}^2+\Delta^2}$ (we also use symbol $E_{l\bd{k}}$ for $\sqrt{\xi_{l\bd{k}}^2+\Delta^2}$).

After the diagonalization, $\ham_\mathcal{C}(\bd{q})$ of Eq. \eqref{eq:hcq} reads
\begin{align}
\ham_\mathcal{C}(\bd{q})=&\sum_\bd{k}\big[\widetilde{\bds{\gamma}}_{\bd{k},\bd{q}}^\dagger\hat{E}_\bd{k}(\bd{q})\widetilde{\bds{\gamma}}_{\bd{k},\bd{q}}+\sum_{l\in\mathcal{C}}\xi^\downarrow_{l,-\bd{k}+\bd{q}}\big]+N\sum_{\alpha=1}^s\frac{|\Delta_{\bd{q},\alpha}|^2}{U_{\alpha}}
\end{align}
where $\widetilde{\bds{\gamma}}_{\bd{k},\bd{q}}$ is the Bogoliubov band spinor that exactly diagonalizes $H_\bd{k}(\bd{q})$. The projected grand potential is
\begin{align}\label{eq:grandc}
\Omega_\mathcal{C}=&-\frac{1}{\beta}\ln\tr\{e^{-\beta\ham_{\mathcal{C}}(\bd{q})}\}\nonumber\\
=&-\frac{1}{\beta}\sum_\bd{k}\tr\ln(1+e^{-\beta\hat{E}_\bd{k}(\bd{q})})+\sum_\bd{k}\sum_{l\in\mathcal{C}}\xi^\downarrow_{l,-\bd{k}+\bd{q}} \nonumber\\
&+N\sum_{\alpha=1}^s\frac{|\Delta_{\bd{q},\alpha}|^2}{U_{\alpha}}
\end{align}  
In $D_{s,\mu\nu}^{(1)}=\partial_\mu\partial_\nu\Omega_\mathcal{C}|_{\bd{q}=0}$, the derivative acts on explicit $\bd{q}$ only and does not act on $\mu_\bd{q}$ or $\hat{\Delta}_\bd{q}$, giving \cite{peotta2015superfluidity}:
\begin{align}\label{eq:d1ofe}
D^{(1)}_{s,\mu\nu}=&-\frac{1}{2}\sum_\bd{k}\tr\bigg\{\frac{\partial^2\hat{E}_\bd{k}(\bd{q})}{\partial q_\mu\partial q_\nu}\tanh\frac{\beta \hat{E}_\bd{k}(\bd{q})}{2} \nonumber\\
&+\frac{\beta}{2}\frac{\partial\hat{E}_\bd{k}(\bd{q})}{\partial q_\mu}\frac{\partial\hat{E}_\bd{k}(\bd{q})}{\partial q_\nu}\text{sech}^2\frac{\beta \hat{E}_\bd{k}(\bd{q})}{2}\bigg\}\bigg|_{\bd{q}=0}.
\end{align}
  
Eq. \eqref{eq:d1ofe} involves the first and second-derivative of $E_{l\bd{k}\pm}(\bd{q})$. If we treat $\bd{q}$ as perturbation and find the eigenvalues of $H_\bd{k}(\bd{q})$ up to the $q^2$ order, then $D_s^{(1)}$ can be determined. To proceed, we approximate $\Lambda_{\bd{k}} (\bd{q})$ as
\begin{align}\label{eq:uck}
\Lambda_{\bd{k}} (\bd{q}) \simeq U_\bd{k}=U_{1\bd{k}}\oplus...\oplus U_{n\bd{k}},
\end{align}
where
\begin{align}
U_{l\bd{k}}=\begin{pmatrix}
	w_{l\bd{k}}&-v_{l\bd{k}}\\
	v_{l\bd{k}}&w_{l\bd{k}}
\end{pmatrix}
\end{align}
is the Bogoliubov transformation for the $l^{th}$ band at $\bd{q}=0$, with
\begin{align}
\label{eq:wv}
w_{l\bd{k}}=\sqrt{\frac{1}{2}\bigg(1+\frac{\xi_{l\bd{k}}}{E_{l\bd{k}}}\bigg)},\,\,\,v_{l\bd{k}}=\sqrt{\frac{1}{2}\bigg(1-\frac{\xi_{l\bd{k}}}{E_{l\bd{k}}}\bigg)}.
\end{align}
Transformation $U_\bd{k}$ makes the off-diagonal elements of $H_\bd{k}(\bd{q})$ small therefore splits into two parts,
\begin{align}\label{eq:twoh}
U_\bd{k}^\dagger H_\bd{k}(\bd{q})U_\bd{k}=\hat{E}_{\bd{k}}(0)+H_{\bd{k}}^{(1)}(\bd{q}),
\end{align}
where $\hat{E}_\bd{k}(0)=\text{diag}\{E_{1\bd{k}},-E_{1\bd{k}},...,E_{n\bd{k}},-E_{n\bd{k}}\}$ is the unperturbed PHS Bogoliubov spectrum, and $H^{(1)}_{\bd{k}}(\bd{q})$ is the perturbing matrix.

This is equivalent to expressing $\ham_{\mathcal{C}}(\bd{q})$ in the Bogoliubov band basis defined by $|l\bd{k}\pm\ra\equiv\gamma^\dagger_{l\bd{k}\pm}(\bd{q})|0\ra$, where $\gamma_{l\bd{k}\pm}$ denote the Bogoliubov operators of the $l^{th}$ band. The 2$n$-component Bogoliubov band spinor $\bds{\gamma}_{\bd{k},\bd{q}}=(\gamma_{1\bd{k}+}(\bd{q}),\,\, \gamma_{1\bd{k}-}(\bd{q}),\,\,...)^T$ is related to the electronic band operator through
\begin{align}
\bds{\gamma}_{\bd{k},\bd{q}}=U_\bd{k}^\dagger (c_{1,\bd{k}+\bd{q}\uparrow},c_{1,-\bd{k}+\bd{q}\downarrow}^\dagger,...)^T.
\end{align}
As we see below, the separation Eq. \eqref{eq:twoh} results in an excitation Hamiltonian
\begin{align}
\ham_{\mathcal{C}}^{(1)}(\bd{q})=\sum_\bd{k}\bds{\gamma}_{\bd{k},\bd{q}}^\dagger H_\bd{k}^{(1)}(\bd{q})\bds{\gamma}_{\bd{k},\bd{q}},
\end{align}
which separately accounts for the SW contributions from intra- and inter-band transitions.

We first assume that at any $\bd{k}$-point, no two bands of the composite are degenerate ($\xi_{l\bd{k}}=\xi_{l'\bd{k}}$), whether accidentally or enforced by symmetry; additionally, we assume no two bands have opposite energies ($\xi_{l\bd{k}}=-\xi_{l' \bd{k}}$). These restrictions mean the $2n$ Bogoliubov bands at $\bd{q}=0$ are disconnected throughout the Brillouin zone, so nondegenerate perturbation can be applied. We find the energy eigenvalues of $H_\bd{k}(\bd{q})$ up to $O(q^2)$ terms give $n$ quasiparticle bands and $n$ quasihole bands
\begin{widetext}
\begin{align}
\label{eq:elq}
&E_{l\bd{k}+}(\bd{q})=E_{l\bd{k}}+H_{\bd{k}}^{(1)}(\bd{q})_{2l-1,2l-1}+\sum_{l'\in\mathcal{C},l'\neq l}\frac{|H_{\bd{k}}^{(1)}(\bd{q})_{2l-1,2l'-1}|^2}{E_{l\bd{k}}-E_{l'\bd{k}}}+\sum_{l'\in\mathcal{C}}\frac{|H_{\bd{k}}^{(1)}(\bd{q})_{2l-1,2l'}|^2}{E_{l\bd{k}}+E_{l'\bd{k}}}+O(q^3), \nonumber\\
&E_{l\bd{k}-}(\bd{q})=-E_{l\bd{k}}+H_{\bd{k}}^{(1)}(\bd{q})_{2l,2l}+\sum_{l'\in\mathcal{C}}\frac{|H_{\bd{k}}^{(1)}(\bd{q})_{2l,2l'-1}|^2}{-E_{l\bd{k}}-E_{l'\bd{k}}}+\sum_{l'\in\mathcal{C},l'\neq l}\frac{|H_{\bd{k}}^{(1)}(\bd{q})_{2l,2l'}|^2}{-E_{l\bd{k}}+E_{l'\bd{k}}}+O(q^3).
\end{align}
\end{widetext}
Here $H_{\bd{k}}^{(1)}(\bd{q})_{ij}$ denotes the $(i,j)^{th}$ entry of matrix $H_{\bd{k}}^{(1)}(\bd{q})$. In terms of the Bogoluibov basis, they are
\begin{align}
\label{eq:Mateledef}
&H_{\bd{k}}^{(1)}(\bd{q})_{2l-1,2l'-1}\equiv\la l\bd{k}+|H_{\bd{k}}^{(1)}(\bd{q})|l'\bd{k}+\ra, \nonumber\\
&H_{\bd{k}}^{(1)}(\bd{q})_{2l,2l'}\equiv\la l\bd{k}-|H_{\bd{k}}^{(1)}(\bd{q})|l'\bd{k}-\ra, \nonumber\\
&H_{\bd{k}}^{(1)}(\bd{q})_{2l-1,2l'}\equiv\la l\bd{k}+|H_{\bd{k}}^{(1)}(\bd{q})|l'\bd{k}-\ra.
\end{align}
which can be explicitly computed from Eq. \eqref{eq:bdgmatrix} and \eqref{eq:uck}.

After some algebra, we find the derivatives of Eq. \eqref{eq:elq} with respect to $\bd{q}$ (which do not act on $\mu_\bd{q}$ or $\hat{\Delta}_\bd{q}$) give
\begin{align}\label{eq:de1}
\partial_\mu E_{l\bd{k}\pm}(\bd{q})\big|_{\bd{q}=0}=\partial_\mu\xi_{l\bd{k}},
\end{align}
\begin{align}\label{eq:de2}
\partial_\mu\partial_\nu E_{l\bd{k}\pm}(\bd{q})&\big|_{\bd{q}=0}=\pm \bigg\{\frac{\xi_{l\bd{k}}}{E_{l\bd{k}}}\partial_\mu\partial_\nu\xi_{l\bd{k}}-\frac{4\Delta^2}{E_{l\bd{k}}}g^l_{\mu\nu}(\bd{k}) \nonumber\\
-\sum_{l'\in\mathcal{C},l'\neq l}8&\Delta^2\bigg[\frac{p^{(+)}_{ll'}(\bd{k})}{E_{l\bd{k}}+E_{l'\bd{k}}}+\frac{p^{(-)}_{ll'}(\bd{k})}{E_{l\bd{k}}-E_{l'\bd{k}}}\bigg]g^{ll'}_{\mu\nu}(\bd{k})\bigg\},
\end{align}
where the coherence factors
\begin{align}\label{eq:cohedef}
&p^{(+)}_{ll'}(\bd{k})=\frac{1}{2}\bigg(1+\frac{\xi_{l\bd{k}}\xi_{l'\bd{k}}+\Delta^2}{E_{l\bd{k}}E_{l'\bd{k}}}\bigg)=(w_{l\bd{k}}w_{l'\bd{k}}+v_{l\bd{k}}v_{l'\bd{k}})^2,\nonumber\\
&p^{(-)}_{ll'}(\bd{k})=\frac{1}{2}\bigg(1-\frac{\xi_{l\bd{k}}\xi_{l'\bd{k}}+\Delta^2}{E_{l\bd{k}}E_{l'\bd{k}}}\bigg)=(w_{l\bd{k}}v_{l'\bd{k}}-v_{l\bd{k}}w_{l'\bd{k}})^2
\end{align}
come from the second derivative $\partial_\mu\partial_\nu|H_{\bd{k}}^{(1)}(\bd{q})_{2l-1,2l'}|^2$, $\partial_\mu\partial_\nu|H_{\bd{k}}^{(1)}(\bd{q})_{2l,2l'-1}|^2$ and $\partial_\mu\partial_\nu|H_{\bd{k}}^{(1)}(\bd{q})_{2l-1,2l'-1}|^2$, $\partial_\mu\partial_\nu|H_{\bd{k}}^{(1)}(\bd{q})_{2l,2l'}|^2$, respectively. Physically, $p^{(+)}_{ll'}(\bd{k})$ accounts for the transitions between a quasiparticle and quasihole band, while $p^{(-)}_{ll'}(\bd{k})$ accounts for transitions between two quasiparticle bands or between two quasihole bands.

Also, one notices that the intra- and inter-band quantum metric comes from the expansion of intra- and inter-band gap functions. If the fluctuation of $\hat{\Delta}_\bd{q}$ is ignored, they are (see definition Eq. \eqref{eq:interbandgap})
\begin{align}\label{eq:deltaexpand1}
\Delta_{l,\bd{k}}(\bd{q})=&\Delta\{1-2\la u_{l\bd{k}}|\partial_\mu u_{l\bd{k}}\ra q_\mu \nonumber\\
&-[\la\partial_\mu u_{l\bd{k}}|\partial_\nu u_{l\bd{k}}\ra+c.c.]q_\mu q_\nu\}+O(q^3),
\end{align}
\begin{align}\label{eq:deltaexpand2}
\Delta_{ll',\bd{k}}(\bd{q})=-2\Delta\la u_{l\bd{k}}|\partial_\mu u_{l'\bd{k}}\ra q_\mu+O(q^2),\,\,\,l\neq l'.
\end{align}
Since all matrix elements $H_{\bd{k}}^{(1)}(\bd{q})_{ij}\sim O(q)$, the intra-band quantum metric $g^l_{\mu\nu}$ in Eq. \eqref{eq:de2} comes from the $O(q^2)$ term of $H_{\bd{k}}^{(1)}(\bd{q})_{2l-1,2l-1}$ combined with the $O(q)$ term of $H_{\bd{k}}^{(1)}(\bd{q})_{2l-1,2l}$ or the $O(q^2)$ term of $H_{\bd{k}}^{(1)}(\bd{q})_{2l,2l}$ combined with the $O(q)$ term of $H_{\bd{k}}^{(1)}(\bd{q})_{2l,2l-1}$. Whereas the inter-band quantum metric $g^{ll'}_{\mu\nu}$ only comes from the $O(q)$ term of $H_{\bd{k}}^{(1)}(\bd{q})_{2l-1,2l'-1}$, $H_{\bd{k}}^{(1)}(\bd{q})_{2l-1,2l'}$, $H_{\bd{k}}^{(1)}(\bd{q})_{2l,2l'-1}$ and $H_{\bd{k}}^{(1)}(\bd{q})_{2l,2l'}$ of $l\neq l'$. The transition processes they represent are illustrated in Fig. \ref{fig:transition}.

Inserting Eq. \eqref{eq:de1}, \eqref{eq:de2} into Eq. \eqref{eq:d1ofe} one gets Eq. \eqref{eq:d1t} in the main text.
\subsection{Degenerate Perturbation}
To get Eq. \eqref{eq:d1t}, we assumed that there are not any two Bogoliubov bands at $\bd{q}=0$ being degenerate at any $\bd{k}$ point, so nondegenerate perturbation was used. Otherwise, if a few quasiparticle bands are degenerate at some isolated $\bd{k}$ points or completely degenerate throughout the momentum space (which requires the band energy $\xi_{l\bd{k}}=\pm\xi_{l'\bd{k}}$), then degenerate perturbation must be applied.

However, using standard degenerate perturbation, one can prove that Eq. \eqref{eq:d1t} remains the same even in the presence of band degeneracy, which agrees with the general Kubo formula result. In calculating current responses using the Kubo formula, whenever two bands become degenerate, the only change is to make the substitution
\begin{align}
\frac{n(E_1)-n(E_2)}{E_1-E_2}\rightarrow \frac{dn(E)}{dE}\bigg|_{E=E_1}.
\end{align}
Similarly, for Eq. \eqref{eq:d1t}, when $E_{l\bd{k}}=E_{l'\bd{k}}$, one only needs to replace
\begin{align}
\frac{\tanh(\beta E_{l\bd{k}}/2)-\tanh(\beta E_{l'\bd{k}}/2)}{E_{l\bd{k}}-E_{l'\bd{k}}}\rightarrow \frac{\beta}{2}\text{sech}^2\frac{\beta E_{l\bd{k}}}{2}.
\end{align}
Therefore, we conclude that band degeneracy poses no additional mathematical difficulty for expressing the SW $D_s^{(1)}$ as Eq. \eqref{eq:d1t}.
\section{Positive Semidefiniteness of the Superfluid Weight}
\label{app:pd}
The purpose of this appendix is to show that the first term $D_s^{(1)}$, Eq. \eqref{eq:d1t} is positive semidefinite. This proof will be independent of lattice geometry. Since the total SW $D_s=D_s^{(1)}+D_s^{(2)}$ is geometry-independent and $D_s^{(2)}$ can be made zero by choosing the MQM, it also shows the total $D_s$ is positive semidefinite for the uniform pairing channel, regardless of temperature and any composite bands.

One first notices that the function
\begin{align}\label{eq:ineq}
\tanh\frac{\beta E_{l\bd{k}}}{2}-\frac{\beta E_{l\bd{k}}}{2}\text{sech}^2\frac{\beta E_{l\bd{k}}}{2}>0,
\end{align}
so the conventional part of Eq. \eqref{eq:d1t} is positive semidefinite. To analyze the geometric term, we break the intra-band quantum metric $g^l_{\mu\nu}$ into inter-band ones $g^{ll'}_{\mu\nu}$, and collect all the terms with the same $g^{ll'}_{\mu\nu}$ (including the symmetric term $l\leftrightarrow l'$), to arrive at the following split expression,
\begin{widetext}
\begin{align}\label{eq:split2}
D^{(1)\text{geo}}_{s,\mu\nu}=&-\sum_\bd{k}\sum_{l\in\mathcal{C},l'\notin\mathcal{C}}\frac{4\Delta^2}{E_{l\bd{k}}}\tanh\frac{\beta E_{l\bd{k}}}{2}g^{ll'}_{\mu\nu}(\bd{k}) \nonumber\\
&-\sum_\bd{k}\sum_{l,l'\in\mathcal{C},l'>l}4\Delta^2\bigg\{\tanh\frac{\beta E_{l\bd{k}}}{2}\bigg[\frac{1}{E_{l\bd{k}}}-\frac{2p_{ll'}^{(+)}(\bd{k})}{E_{l\bd{k}}+E_{l'\bd{k}}}-\frac{2p_{ll'}^{(-)}(\bd{k})}{E_{l\bd{k}}-E_{l'\bd{k}}}\bigg]+l\leftrightarrow l'\bigg\}g^{ll'}_{\mu\nu}(\bd{k}).
\end{align}
Since each $g^{ll'}_{\mu\nu}$ ($l\neq l'$) is negative semidefinite, all we need to show is that each quantity
\begin{align}\label{eq:fllp}
f(\xi_{l\bd{k}},\xi_{l'\bd{k}})=\tanh\frac{\beta E_{l\bd{k}}}{2}\bigg[\frac{1}{E_{l\bd{k}}}-\frac{2p_{ll'}^{(+)}(\bd{k})}{E_{l\bd{k}}+E_{l'\bd{k}}}-\frac{2p_{ll'}^{(-)}(\bd{k})}{E_{l\bd{k}}-E_{l'\bd{k}}}\bigg]+l\leftrightarrow l'
\end{align}
is positive. Using definition \eqref{eq:cohedef}, we find
\begin{align}
f(\xi_{l\bd{k}},&\xi_{l'\bd{k}})=-\bigg(\frac{1}{E_{l\bd{k}}}\tanh\frac{\beta E_{l\bd{k}}}{2}-\frac{1}{E_{l'\bd{k}}}\tanh\frac{\beta E_{l'\bd{k}}}{2}\bigg)\frac{\xi_{l\bd{k}}-\xi_{l'\bd{k}}}{\xi_{l\bd{k}}+\xi_{l'\bd{k}}}.
\end{align}
Note the factor $\frac{\xi_{l\bd{k}}-\xi_{l'\bd{k}}}{\xi_{l\bd{k}}+\xi_{l'\bd{k}}}=\frac{(\xi_{l\bd{k}}-\xi_{l'\bd{k}})^2}{(E_{l\bd{k}}+E_{l'\bd{k}})(E_{l\bd{k}}-E_{l'\bd{k}})}$, and function $(1/x)\tanh x$ has negative slope at $x>0$ (same as Eq. \eqref{eq:ineq}), therefore $f(\xi_{l\bd{k}},\xi_{l'\bd{k}})>0$ and the positive semidefiniteness is proved.
\section{Calculation of the Pairing Amplitude}
\label{app:amplitude}
In this appendix, we give details of using the perturbation method to calculate the pairing amplitude $\la c_{l',-\bd{k}+\bd{q}\downarrow}c_{l,\bd{k}+\bd{q}\uparrow}\ra_\bd{q}$.

Recall that under transformation $U_\bd{k}$, the mean-field Hamiltonian can be expressed as
\begin{align}
\ham_{\mathcal{C}}(\bd{q})=\sum_\bd{k}\bds{\gamma}^\dagger_{\bd{k},\bd{q}}[\hat{E}_{\bd{k}}(0)+H_{\bd{k}}^{(1)}(\bd{q})]\bds{\gamma}_{\bd{k},\bd{q}},
\end{align}
where the perturbation matrix $H_{\bd{k}}^{(1)}$ mixes Bogoliubov bands of different indices. Using the standard perturbation method, the annihilation operator of the band-resolved quasiparticles that exactly diagonalize $\ham_{\mathcal{C}}(\bd{q})$, $\widetilde{\gamma}$, can be expressed as a linear combination of $ \gamma$ operators:
\begin{align}\label{eq:statepm}
&\widetilde{\gamma}_{l\bd{k}+}(\bd{q})=\gamma_{l\bd{k}+}(\bd{q})+\sum_{l'\in\mathcal{C},l'\neq l}\frac{H_{\bd{k}}^{(1)}(\bd{q})_{2l-1,2l'-1}}{E_{l\bd{k}}-E_{l'\bd{k}}}\gamma_{l'\bd{k}+}(\bd{q})+\sum_{l'\in\mathcal{C}}\frac{H_{\bd{k}}^{(1)}(\bd{q})_{2l-1,2l'}}{E_{l\bd{k}}+E_{l'\bd{k}}}\gamma_{l'\bd{k}-}(\bd{q})+O(q^2), \nonumber\\
&\widetilde{\gamma}_{l\bd{k}-}(\bd{q})=\gamma_{l\bd{k}-}(\bd{q})+\sum_{l'\in\mathcal{C}}\frac{H_{\bd{k}}^{(1)}(\bd{q})_{2l,2l'-1}}{-E_{l\bd{k}}-E_{l'\bd{k}}}\gamma_{l'\bd{k}+}(\bd{q})+\sum_{l'\in\mathcal{C},l'\neq l}\frac{H_{\bd{k}}^{(1)}(\bd{q})_{2l,2l'}}{-E_{l\bd{k}}+E_{l'\bd{k}}}\gamma_{l'\bd{k}-}(\bd{q})+O(q^2).
\end{align}
\end{widetext}
Here, to expend $H_{\bd{k}}^{(1)}(\bd{q})_{ij}$ in powers of $\bd{q}$, we must take into account the $\bd{q}$-dependence of $\hat{\Delta}_\bd{q}$, therefore use
\begin{align}
\Delta_{ll',\bd{k}}(\bd{q})=&\Delta\delta_{ll'}+(-2\Delta\la u_{l\bd{k}}|\partial_\mu u_{l'\bd{k}}\ra \nonumber\\
&+\sum_{\alpha=1}^s u_{l\bd{k}\alpha}^*u_{l'\bd{k}\alpha}\derd_\mu \Delta_\alpha)q_\mu+O(q^2),
\end{align}
instead of Eq. \eqref{eq:deltaexpand1}, \eqref{eq:deltaexpand2}. The normalization of $\widetilde{\gamma}$ operators is ignored as it leads to $O(q^2)$ corrections. Eq. \eqref{eq:statepm} can be viewed as a linear transformation $\widetilde{\bds{\gamma}}=A\bds{\gamma}$ between the spinor $\widetilde{\bds{\gamma}}$ and $\bds{\gamma}$, where $A_{ij}=\delta_{ij}+a_{ij}$ is a matrix close to identity, with $a_{ij}\sim O(q)$. The inverse of $A$ has an approximate form $(A^{-1})_{ij}=\delta_{ij}-a_{ij}+O(q^2)$, which enables us to write operators $\gamma$ in terms of $\widetilde{\gamma}$. Since the electron operators $c$ are related to $\gamma$ through the transformation $ U_{\bd{k}}$, we can finally write $c$ in terms of $\widetilde{\gamma}$.

Since the $\widetilde{\gamma}$ operators exactly diagonalize the BdG Hamiltonian of composite bands, we have
\begin{align}\label{eq:diagonalize}
\la\widetilde{\gamma}_{l\bd{k}\pm}(\bd{q})^\dagger\widetilde{\gamma}_{l\bd{k}\pm}(\bd{q})\ra_\bd{q}\approx n_F(\pm E_{l\bd{k}}),
\end{align}
to the lowest order of $\bd{q}$, where $n_F$ is the Fermi-Dirac function. Finally, one obtains the pairing amplitude
\begin{widetext}
\begin{align}
\la c_{l',-\bd{k}+\bd{q}\downarrow}c_{l,\bd{k}+\bd{q}\uparrow}\ra_\bd{q}=\begin{cases}
\displaystyle
-\frac{\tanh(\beta E_{l\bd{k}}/2)}{2E_{l\bd{k}}}\bigg[\Delta+\big(-2\Delta\la u_{l\bd{k}}|\partial_\mu u_{l\bd{k}}\ra+\sum_{\beta=1}^s|u_{l\bd{k}\beta}|^2\derd_\mu \Delta_\beta\big)q_\mu\bigg]+O(q^2),& l=l',\\
\displaystyle
\frac{1}{2}\bigg[\frac{p^{(+)}_{ll'}(\bd{k})}{E_{l\bd{k}}+E_{l'\bd{k}}}\bigg(\tanh\frac{\beta E_{l\bd{k}}}{2}+\tanh\frac{\beta E_{l'\bd{k}}}{2}\bigg)+\frac{p^{(-)}_{ll'}(\bd{k})}{E_{l\bd{k}}-E_{l'\bd{k}}}\bigg(\tanh\frac{\beta E_{l\bd{k}}}{2}-\tanh\frac{\beta E_{l'\bd{k}}}{2}\bigg)\bigg]\\
\displaystyle
\times\big(2\Delta\la u_{l\bd{k}}|\partial_\mu u_{l'\bd{k}}\ra-\sum_{\beta=1}^s u_{l\bd{k}\beta}^* u_{l'\bd{k}\beta}\derd_\mu \Delta_\beta\big)q_\mu+O(q^2),& l\neq l'.
\end{cases}
\end{align}
\end{widetext}
From this to get Eq. \eqref{eq:mmatrix} and \eqref{eq:vvector}, we have also used the $\bd{q}=0$ gap equation
\begin{align}
1=\frac{U_\alpha}{N}\sum_{\bd{k}}\sum_{l\in\mathcal{C}}|u_{l\bd{k}\alpha}|^2\frac{\tanh(\beta E_{l\bd{k}}/2)}{2E_{l\bd{k}}},\,\,\,\forall\,\alpha.
\end{align}
\section{Positive Semidefiniteness and Rank of the Hessian Matrix for Uniform Pairing}
\label{app:rank}
This appendix serves to prove the rank and positive semidefiniteness of the Hessian matrix $M_{\alpha\beta}$ in Eq. \eqref{eq:mmatrix}.	 

We consider a general class of $s$-dim real symmetric matrices $M_{\alpha\beta}$, which have a kernel eigenvector $\bd{v}_0=(1,1,...,1)^T$. It implies $\sum_{\beta=1}^sM_{\alpha\beta}=\sum_{\alpha=1}^sM_{\alpha\beta}=0$. With this property, we write $M_{\alpha\beta}$ as
\begin{align}\label{eq:mform}
M_{\alpha\beta}=\begin{cases}
\sum_{\gamma=1,\gamma\neq\alpha}^s C_{\alpha\gamma},&\alpha=\beta\\
-C_{\alpha\beta},&\alpha\neq\beta
\end{cases}
\end{align}
where $C_{\alpha\beta}$ ($\alpha\neq \beta$) are just the off-diagonal elements of $M$ with a minus sign. We give three propositions below.

\begin{propt}\label{prop:cpositive}
\normalfont
If $C_{\alpha\beta}>0$ for all $1\leqslant\alpha,\beta\leqslant s,\,\alpha\neq\beta$, then $M_{\alpha\beta}$ is positive semidefinite and $\text{Rank}(M)=s-1$.
\end{propt}

We use the criterion that all its leading principal minors are nonnegative to show the positive semidefiniteness. Let's denote the leading principal minor of $M$ of order $k$ ($1\leqslant k\leqslant s$) by $D_k(M)$, we find
\begin{align}\label{eq:d1m}
D_1(M)=M_{11}=\sum_{\alpha\neq 1}^sC_{1\alpha}\ges 0,
\end{align}
\begin{align}\label{eq:d2m}
D_2(M)=\det\begin{pmatrix}
M_{11}&M_{12}\\
M_{21}&M_{22}
\end{pmatrix}=\sum_{\alpha_1\neq 1,\alpha_2\neq 2}^sC_{1\alpha_1}C_{2\alpha_2}-C_{12}C_{21},
\end{align}
and
\begin{align}\label{eq:dkm}
D_k(M)=\sideset{}{'}\sum_{\alpha_1,...,\alpha_k}^s C_{1\alpha_1}C_{2\alpha_2}...C_{k\alpha_k}.
\end{align}
Here, $\sum'$ means the summation imposed by the single rule that there is not any subset $\{i_1,...,i_j\}\subseteq \{1,...,k\}$ such that $\{\alpha_{i_1},...,\alpha_{i_j}\}$ is its permutation. This single rule leads to the following sub-rules: (i) no $C_{\alpha\alpha}$ appears in the sum (e.g., \eqref{eq:d1m}), since $\{\alpha\}$ is a permutation of $\{\alpha\}$; (ii) no squares of any $C$ appears in the sum, i.e. for $\alpha\neq \beta$, $C_{\alpha\beta}C_{\beta\alpha}$ is not allowed since $\{\beta,\alpha\}$ is a permutation of $\{\alpha,\beta\}$ (e.g., \eqref{eq:d2m}); (iii) in every term of $D_k(M)$, there is at least an $\alpha_j$ ($1\les j\les k$) that is not in the set $\{1,...,k\}$. Sub-rule (iii) is a simple result of the general permutation rule.

Using (iii), it is easy to show that $\det M=D_s(M)=0$, thus $\text{Rank}(M)\les s-1$. Moreover, since each leading principal minor $D_k(M)\ges 0$, $M_{\alpha\beta}$ is positive semidefinite.

Next, we prove $\text{Rank}(M)=s-1$ by contradiction---if $\text{Rank}(M)\les s-2$, then any minor of order $s-1$ should vanish. However,
\begin{align}\label{eq:dnminus1}
D_{s-1}(M)=\sideset{}{'}\sum_{\alpha_1,...,\alpha_{s-1}}^s C_{1\alpha_1}C_{2\alpha_2}...C_{s-1,\alpha_{s-1}}>0
\end{align}
since it contains at least one term $C_{1s}C_{2s}...C_{s-1,s}$, a contradiction (QED).

We want to emphasize that Prop. \ref{prop:cpositive} together with Eq. \eqref{eq:mform} is a statement about UPC. It can be equivalently stated as ``UPC leads to a positive semidefinite $M$ and $\text{Rank}(M)=s-1$". We cannot prove the most general case, but will prove two special cases below, which still contain enough information for understanding this general property.
\begin{propt}\label{prop:singleband}
\normalfont
Under the UPC, if an isolated band $m$ contains $s$ orbitals, then the Hessian matrix $M$ is positive semidefinite and $\text{Rank}(M)=s-1$.
\end{propt}

The condition above can be stated as ``for each $1\les\alpha\les s$ there exists a neighborhood of some $\bd{k}$-point in Brillouin zone such that $|u_{m\bd{k}\alpha}|\neq 0$". Based on Prop. \ref{prop:cpositive}, we only need to show $C_{\alpha\beta}>0$. From Eq. \eqref{eq:mmatrix} we find for a single isolated band $m$,
\begin{align}
C_{\alpha\beta}=\sum_\bd{k}\frac{\tanh(\beta E_{m\bd{k}}/2)}{2E_{m\bd{k}}}|u_{m\bd{k}\alpha}|^2|u_{m\bd{k}\beta}|^2,\,\,\,\alpha\neq\beta,
\end{align}
which is positive. Therefore, the proposition is proved.

By induction, we know that whenever an orbital is removed from the band, one positive eigenvalue of $M$ will approach 0, lowering the rank by 1. This can be easily seen from that $|u_{m\bd{k}r}|\rightarrow 0$ if the $r^{th}$ orbital is removed. Then $M$ will have an additional kernel eigenvector $\bd{e}_r=(...,0,1,0,...)^T$ (with its $r^{th}$ component nonzero).
\begin{propt}\label{prop:composite}
\normalfont
Under the UPC, for a composite of $n$ bands formed from $s$ orbitals, if (i) the composite is incomplete ($n<s$), (ii) the $n$ bands are energetically close to each other compared to the interaction scale $\Delta$, and (iii) the composite contains all the $s$ orbitals, then the Hessian matrix $M$ is positive semidefinite and $\text{Rank}(M)=s-1$.
\end{propt}

This proposition is based on three conditions, but the first two can be relaxed. Similarly, the third condition can be stated as ``for each $1\les\alpha\les s$ there exists some band $l\in\mathcal{C}$ and a neighborhood of some $\bd{k}$-point in Brillouin zone such that $|u_{l\bd{k}\alpha}|\neq 0$". The second condition states that the $n$ bands form an ICFB. With this condition one can omit the band index in $E_{l\bd{k}}$, $E_{l\bd{k}}\rightarrow E_\bd{k}$. From Eq. \eqref{eq:mmatrix}, $C_{\alpha\beta}$ of the composite bands can be expressed as
\begin{align}\label{eq:ccomposite}
C_{\alpha\beta}&\approx\sum_\bd{k}\frac{\tanh(\beta E_{\bd{k}}/2)}{2E_{\bd{k}}}\sum_{l,l'\in\mathcal{C}}u_{l'\bd{k}\alpha}^*u_{l\bd{k}\alpha}u_{l\bd{k}\beta}^*u_{l'\bd{k}\beta} \nonumber\\
&=\sum_\bd{k}\frac{\tanh(\beta E_{\bd{k}}/2)}{2E_{\bd{k}}}|b_{\bd{k},\alpha\beta}|^2,
\end{align}
where we have defined
\begin{align}
b_{\bd{k},\alpha\beta}=\sum_{l\in\mathcal{C}} u_{l\bd{k}\alpha}^*u_{l\bd{k}\beta}.
\end{align}
Eq. \eqref{eq:ccomposite} is the leading term of $C_{\alpha\beta}$ if it is expanded in powers of $\xi_{l\bd{k}}-\xi_{l'\bd{k}}$. Since it is required that the composite is incomplete ($n<s$), we have $|b_{\bd{k},\alpha\beta}|$ strictly greater than 0. Then $C_{\alpha\beta}>0$, and the proposition is proved. If instead $n=s$, then $b_{\bd{k},\alpha\beta}=\delta_{\alpha\beta}$, making the leading term of $C_{\alpha\beta}$ vanish.

For cases other than those stated in Prop. \ref{prop:composite}, one should be able to numerically check that $C_{\alpha\beta}>0$ always holds under UPC. Analogous to the single isolated band case, whenever an $r^{th}$ orbital is removed from the composite, $\bd{e}_r$ becomes an additional kernel eigenvector and $\text{Rank}(M)$ is lowered by 1.
\section{Equivalence of the Two Definitions of Minimal Quantum Metric}
\label{app:equivalence}
In this appendix, we show how the two definitions of MQM given in Sec. \ref{sec:mqm} are equivalent to each other based on the rank information of matrix $\partial^2\Omega/\partial\Delta^I_\alpha\partial\Delta^I_\beta$.

Consider the geometric transformation that each orbital's position transforms as $\bd{x}_\alpha=\bd{x}^0_\alpha+\delta \bd{x}_\alpha$. During this transformation, hopping integrals $t^\sigma_{ij,\alpha\beta}$ are fixed, while the Bloch components $u_{l\bd{k}\alpha}$ undergo the change
\begin{align}\label{eq:deltau}
u_{l\bd{k}\alpha}=e^{-i\bd{k}\cdot\delta\bd{x}_\alpha}u^0_{l\bd{k}\alpha}.
\end{align}
The intra- and inter-band quantum metric transform accordingly, $g^l_{\mu\nu}=g^{l,0}_{\mu\nu}+\delta g^l_{\mu\nu}$, $g^{ll'}_{\mu\nu}=g^{ll',0}_{\mu\nu}+\delta g^{ll'}_{\mu\nu}$, with
\begin{widetext}
\begin{align}\label{eq:deltag}
&\delta g^l_{\mu\nu}(\bd{k})=\la \delta x_{\mu}\delta x_{\nu}\ra^0_{l\bd{k}}-\la \delta x_{\mu}\ra^0_{l\bd{k}} \la \delta x_{\nu}\ra^0_{l\bd{k}} +\big[\big(\frac{i}{2}\sum_\alpha \delta x_{\alpha,\mu}u_{l\bd{k}\alpha}^{0*}\partial_\nu u_{l\bd{k}\alpha}^0+c.c.-i\la \delta x_{\mu}\ra^0_{l\bd{k}}\la u^0_{l\bd{k}}|\partial_\nu u^0_{l\bd{k}}\ra\big)+\mu\leftrightarrow\nu\big], \nonumber\\
&\delta g^{ll'}_{\mu\nu}(\bd{k})=\big[-\frac{1}{2}\la\delta x_\mu\ra_{ll'\bd{k}}^0\la\delta x_\nu\ra_{l'l\bd{k}}^0-\frac{i}{2}\big(\la\delta x_\mu\ra_{ll'\bd{k}}^0\la u_{l'\bd{k}}^0|\partial_\nu u_{l\bd{k}}^0\ra+\la\delta x_\nu\ra_{l'l\bd{k}}^0\la u_{l\bd{k}}^0|\partial_\mu u_{l'\bd{k}}^0\ra\big)\big]+\mu\leftrightarrow\nu.
\end{align}
\end{widetext}
Here $\la\ra^0_{l\bd{k}}$ represents the average over different orbitals in the initial state $u^0_{l\bd{k}}$, while $\la\ra^0_{ll'\bd{k}}$ represents the overlap between different bands, i.e.,
\begin{align}
&\la \delta x_{\mu}\ra^0_{l\bd{k}}=\sum_{\alpha=1}^s \delta x_{\alpha,\mu}|u^0_{l\bd{k}\alpha}|^2, \nonumber\\
&\la \delta x_{\mu}\delta x_{\nu}\ra^0_{l\bd{k}}=\sum_{\alpha=1}^s \delta x_{\alpha,\mu}\delta x_{\alpha,\nu}|u^0_{l\bd{k}\alpha}|^2, \nonumber\\
&\la\delta x_\mu\ra_{ll'\bd{k}}^0=\sum_{\alpha=1}^s \delta x_{\mu,\alpha} u_{l\bd{k}\alpha}^{0*}u_{l'\bd{k}\alpha}^0.
\end{align}

With these established, let's vary the functional $I$ in Eq. \eqref{eq:functional2} around some fixed positions $\{\bd{x}_\alpha^0\}$. One can readily show the following identity,
\begin{align}\label{eq:variationeq}
\frac{\partial I}{\partial x_{\alpha,\mu}}\bigg|_{\{\bd{x}_\alpha^0\}}=\bigg(\frac{2i}{\Delta}\bigg)V^0_{\alpha,\mu},
\end{align}
where $V_{\alpha,\mu}^0$ is Eq. \eqref{eq:vvector} evaluated at $\{\bd{x}_\alpha^0\}$. Based on Definition 2, if $\{\bd{x}_\alpha^0\}$ extremize $I$, then $V^0_{\alpha,\mu}=M_{\alpha\beta}\derd_\mu\Delta_\beta^0=0$, implying $D_s^{(2)}=0$ in Eq. \eqref{eq:fullSW}. Therefore we have proved that Definition 2 leads to Definition 1.

To show the other way requires the rank of the Hessian matrix $\partial^2\Omega/\partial_\mu\Delta_\alpha\partial_\nu\Delta_\beta$ to be $s-1$. The uniform pairing case has been proved in App. \ref{app:rank}. Assuming $\text{Rank}(M)=s-1$, then the vanishing of $D_s^{(2)}$ in Eq. \eqref{eq:fullSW} has two solutions: either (i) $\derd_\mu \Delta_\alpha=0$ or (ii) $\derd_\mu \Delta_1=\derd_\mu \Delta_2=...=\derd_\mu \Delta_s=i\lambda_\mu$, $\mu=x,y$. These two solutions are identical up to a $U(1)$ gauge transformation \cite{chan2022pairing,huhtinen2022revisiting}. To see this, we write the second solution as
\begin{align}
\Delta_{\bd{q},\alpha}=\Delta[1+i\bds{\lambda}\cdot\bd{q}+S_\alpha(q^2)],
\end{align}
where $\bds{\lambda}=(\lambda_x,\lambda_y)$ is a real vector and $S_\alpha(q^2)$ is a complex function of $q^2$ order. A new order parameter that has an overall phase difference can be defined,
\begin{align}
\widetilde{\Delta}_{\bd{q},\alpha}=e^{-i\bds{\lambda}\cdot\bd{q}}\Delta_{\bd{q},\alpha},
\end{align}
which satisfies $\derd_\mu \widetilde{\Delta}_\alpha=0$ and recovers the first solution. Both solutions above lead to the vanishing of extremization Eq. \eqref{eq:variationeq}, since $V_{\alpha,\mu}=M_{\alpha\beta}\derd_\mu\Delta_\alpha$. We have shown that Definition 1 implies Definition 2. Therefore, the two definitions are equivalent.

With the knowledge that $\text{Rank}(M)=s-1$, one can further show the minimal positions exist and are unique, up to fixing the position of one orbital \cite{huhtinen2022revisiting}.

Starting from some general positions $\{\bd{x}_\alpha^0\}$, let's consider the transformation $\bd{x}_\alpha=\bd{x}^0_\alpha+\delta \bd{x}_\alpha$ again, under which $\derd_\mu\Delta_\alpha=\derd_\mu\Delta_\alpha^0+\delta \derd_\mu\Delta_\alpha$ and $V_{\alpha,\mu}=V^0_{\alpha,\mu}+\delta V_{\alpha,\mu}$. Using Eq. \eqref{eq:vvector} and \eqref{eq:deltau}, the change of $V_{\alpha,\mu}$ is found to be
\begin{align}\label{eq:vchange}
\delta V_{\alpha,\mu}=-2i\Delta M_{\alpha\beta}\delta x_{\beta,\mu}.
\end{align}
By the invertibility of $M_{\alpha\beta}$ in the $(s-1)$-dim subspace, we obtain
\begin{align}\label{eq:xchange}
\delta \derd_\mu\Delta_\alpha=-2i\Delta\delta x_{\alpha,\mu}.
\end{align}
Imposing $\derd_\mu\Delta_\alpha^0+\delta \derd_\mu\Delta_\alpha=0$, thus the minimal positions are located at
\begin{align}
\delta x_{\alpha,\mu}=-\frac{i}{2\Delta}\derd_\mu\Delta_\alpha^0,
\end{align}
from the initial positions $\{\bd{x}_{\alpha}^0\}$.

As discussed at the end of Sec. \ref{sec:mqm}, when some orbitals are removed from the composite bands, their positions become irrelevant to the geometric functional $I$. Thus, the MQM orbital positions are hyperlines or hyperplanes in the geometric space. We use two-orbital models as an example to illustrate this in App. \ref{app:2orbital}.
\section{Two-orbital Models}
\label{app:2orbital}
Inserting Eq. \eqref{eq:deltag} into Eq. \eqref{eq:d1t}, and inserting Eq. \eqref{eq:xchange} into Eq. \eqref{eq:change2}, we find both $\delta D_s^{(1)}$ and $\delta D_s^{(2)}$ have a quadratic form of the orbital coordinate change $\delta x_{\alpha,\mu}$, therefore take a parabolic shape in the $3(s-1)$-dim geometric space. The eigenvalues of Hessian matrix $M_{\alpha\beta}$ show the steepness of the parabola, while the eigenvectors tell the directions where the MQM functional undergoes the steepest change. Whenever an orbital is removed from the composite, one eigenvalue of $M_{\alpha\beta}$ reaches zero, and its position becomes irrelevant to the functional $I$. In this appendix, we illustrate this using two-orbital models.

For two-orbital models, one can define the composite to contain either one (valence or conduction) band or both bands, depending on the Fermi energy and interaction scale. Since the position of one orbital can permanently be fixed, we write $I$ as a function $I(\bd{r})$, where $\bd{r}=(x,y)\equiv \bd{x}_2-\bd{x}_1$ is orbital 2's position relative to orbital 1. To calculate $I(\bd{r})$, one can use the method of the Bloch function, i.e., using Eq. \eqref{eq:deltau} to get \eqref{eq:deltag}. Here, instead, we use the Bloch Hamiltonian method.

We choose the lattice geometry where the two orbitals sit on the same site as the reference geometry, $\bd{r}=0$. This defines the Bloch Hamiltonian of a two-orbital model
\begin{align}
h(\bd{k})=d_0(\bd{k})\sigma_0+\sum_{j=x,y,z}d_j(\bd{k})\sigma_j.
\end{align}
(we omitted the spin index, so $h(\bd{k})$ is for spin-$\uparrow$ only and the spin-$\downarrow$ sector is its time-reversal counterpart) where $\sigma_j$ are Pauli matrices in the orbital space. Then, the quantum metric is
\begin{align}\label{eq:goftwoband}
g_{\mu\nu}\equiv g^{v}_{\mu\nu}=g^{c}_{\mu\nu}=-g^{vc}_{\mu\nu}=\frac{1}{4}\partial_\mu\hat{\bd{d}}\cdot \partial_\nu\hat{\bd{d}},
\end{align}
where $\bd{d}=(d_x,d_y,d_z)$ and $\hat{\bd{d}}=\bd{d}/|\bd{d}|$ is a unit vector.

Under the geometric transformation, $h(\bd{k})$ is transformed by the unitary matrix $\mathcal{U}(\bd{k})=\text{diag}\{1,e^{i\bd{k}\cdot\bd{r}}\}$:
\begin{align}
\widetilde{h}(\bd{k})=\mathcal{U}(\bd{k})^\dagger h(\bd{k})\mathcal{U}(\bd{k}),
\end{align}
therefore vector $\bd{d}$ is rotated in the way
\begin{align}
&\widetilde{d}_x=d_x\cos(\bd{k}\cdot\bd{r})+d_y\sin(\bd{k}\cdot\bd{r}), \nonumber\\
&\widetilde{d}_y=-d_x\sin(\bd{k}\cdot\bd{r})+d_y\cos(\bd{k}\cdot\bd{r}), \nonumber\\
&\widetilde{d}_z=d_z,
\end{align}
which gives the quantum metric change
\begin{align}
\widetilde{g}_{\mu\nu}=&g_{\mu\nu}+\frac{1}{4}[(-\hat{d}_x\partial_\mu \hat{d}_y+\hat{d}_y\partial_\mu \hat{d}_x)r_\nu \nonumber\\
&+(-\hat{d}_x\partial_\nu \hat{d}_y+\hat{d}_y\partial_\nu \hat{d}_x)r_\mu+(\hat{d}_x^2+\hat{d}_y^2)r_\mu r_\nu]
\end{align}
where $\mu$ or $r_\mu=x$ or $y$. Then $I(\bd{r})$ of Eq. \eqref{eq:functional2} is
\begin{align}\label{eq:fquadratic1}
I(\bd{r})=I_0+I_{1x}x+I_{1y}y+I_2(x^2+y^2),
\end{align}
with
\begin{align}\label{eq:threef}
&I_0=\sum_\bd{k}f(\bd{k})\tr g(\bd{k}), \nonumber\\
&I_{1\mu}=\frac{1}{2}\sum_\bd{k}f(\bd{k})(-\hat{d}_x\partial_\mu \hat{d}_y+\hat{d}_y\partial_\mu \hat{d}_x), \nonumber\\
&I_2=\frac{1}{4}\sum_\bd{k}f(\bd{k})(\hat{d}_x^2+\hat{d}_y^2).
\end{align}
Here, the form of $f(\bd{k})$ depends on whether the composite contains only one band or both bands. For the former, $f(\bd{k})=f_{v(c)}(\bd{k})=\tanh(\beta E_{v(c)\bd{k}}/2)/E_{v(c)\bd{k}}$; for the latter,
\begin{align}\label{eq:fcomp}
f(\bd{k})=&\frac{1}{E_{v\bd{k}}}\tanh\frac{\beta E_{v\bd{k}}}{2}+\frac{1}{E_{c\bd{k}}}\tanh\frac{\beta E_{c\bd{k}}}{2} \nonumber\\
&-\tanh\frac{\beta E_{v\bd{k}}}{2}\bigg[\frac{2p_{vc}^{(+)}(\bd{k})}{E_{v\bd{k}}+E_{c\bd{k}}}+\frac{2p_{vc}^{(-)}(\bd{k})}{E_{v\bd{k}}-E_{c\bd{k}}}\bigg] \nonumber\\
&-\tanh\frac{\beta E_{c\bd{k}}}{2}\bigg[\frac{2p_{vc}^{(+)}(\bd{k})}{E_{v\bd{k}}+E_{c\bd{k}}}+\frac{2p_{vc}^{(-)}(\bd{k})}{E_{c\bd{k}}-E_{v\bd{k}}}\bigg].
\end{align}

Notice that $I_0,I_{1\mu},I_2$ are all geometry-independent quantities. Completing the squares in Eq. \eqref{eq:fquadratic1} one get
\begin{align}\label{eq:fquadratic2}
I(\bd{r})=I_0-\frac{I_{1x}^2+I_{1y}^2}{4I_2}+I_2\bigg[\bigg(x+\frac{I_{1x}}{2I_2}\bigg)^2+\bigg(y+\frac{I_{1y}}{2I_2}\bigg)^2\bigg],
\end{align}
indicating that the extremal position is located at $\bd{r}_{min}=-(I_{1x}, I_{1y})/(2I_2)$. The sign of the parabola is determined by $\sgn(I_2)=\sgn(f(\bd{k}))$. It turns out that Eq. \eqref{eq:fcomp} is the same as Eq. \eqref{eq:fllp}, which we have proved to be positive. Therefore, for two-orbital models, no matter whether the composite contains one band or both the valence and conduction band, $I_2>0$, and the function is minimized at the point $\bd{r}_{min}$. This is another example showing the Hessian matrix $M$ for a composite of two bands of a two-orbital model is positive semidefinite and $\text{Rank}(M)=1$, with the first two conditions stated in Prop. \ref{prop:composite} relaxed.

Next, we consider two prototype models, the Bravais-lattice BHZ model \cite{bhz2006,qwz2006} and the non-Bravais-lattice Haldane model \cite{haldane1988} to show how the parabola evolves under topological phase transitions. 

The BHZ model \cite{bhz2006,qwz2006} is defined on a square lattice. Choosing the geometric gauge where both orbitals sit on the same site, the Hamiltonian has
\begin{align}
\bd{d}(\bd{k})=(t\sin k_x,t\sin k_y,m+t\cos k_x+t\cos k_y).
\end{align}
Inserting this into Eq. \eqref{eq:threef}, one gets $\bd{r}_{min}=(0,0)$, regardless of the composite, temperature and value of $m/t$. This is enforced by the $C_4$ symmetry of the model hopping graph.

Similarly, the Bloch Hamiltonian of Haldane model has
\begin{align}
&d_0(\bd{k})=2t_2\cos\phi\sum_{i=1}^3\cos(\bd{k}\cdot\bd{a}_i), \nonumber\\
&d_x(\bd{k})=t_1[1+\cos(\bd{k}\cdot\bd{a}_2)+\cos(\bd{k}\cdot\bd{a}_3)], \nonumber\\
&d_y(\bd{k})=t_1[\sin(\bd{k}\cdot\bd{a}_2)-\sin(\bd{k}\cdot\bd{a}_3)], \nonumber\\
&d_z(\bd{k})=m-2t_2\sin\phi\sum_{i=1}^3\sin(\bd{k}\cdot\bd{a}_i).
\end{align}
Inserting this into Eq. \eqref{eq:threef} we find $\bd{r}_{min}=(0,\sqrt{3})$, regardless of the composite, temperature, $t_1,t_2$ and $m$ values. This is enforced by the $C_6$ symmetry of the hopping graph.

As $m/t\rightarrow\infty$, both the BHZ and Haldane model transit to the topologically trivial phase, in which $\hat{d}_x^2+\hat{d}_y^2\propto t^2/m^2\rightarrow 0$ (during the transition we assume $t,t_1,t_2$ are fixed but $m$ is varied). Suppose the composite contains the isolated valence or conduction band only ($\Delta\sim t\ll m$). In that case, we find $f(\bd{k})\sim1/\Delta$ which has no dependence on $m$, therefore $I_2\propto f(\bd{k})(\hat{d}_x^2+\hat{d}_y^2)\rightarrow 0$, implying that the parabola in Eq. \eqref{eq:fquadratic2} becomes flat. This means the functional $I(\bd{r})$ becomes a constant in the geometric space as one orbital is removed from the band.

Conversely, if the composite contains both the valence and conduction band ($\Delta\sim m\gg t$), it always contains two orbitals, even in the topologically trivial phase. In the atomic limit, from Eq. \eqref{eq:fcomp}, one can show
\begin{align}
f(\bd{k})\sim \frac{m^2}{\Delta^3}.
\end{align}
Therefore, $I_2\sim t^2/\Delta^3$ remains finite, and the rank of the Hessian matrix does not decrease. This means $I(\bd{r})$ of Eq. \eqref{eq:fquadratic2} remains parabolic in the atomic limit.
\section{Hessian Matrix of Nonuniform Pairing}
\label{app:npd}
This appendix studies the properties of the matrix of second derivative of grand potential with order parameters beyond the uniform pairing. In particular, we show that this matrix may not be positive semidefinite.

We restrict the interaction to be onsite intra-orbital density-density interaction. We also assume a time-reversal invariant channel $\hat{\Delta}$ in orbital space, i.e., under TRS, the matrix $\hat{\Delta}$ is mapped to itself up to a $U(1)$ phase. Then for our interaction type, the pairing matrix at $\bd{q}=0$ takes the diagonal form $\hat{\Delta}=\text{diag}\{\Delta_1,\Delta_2,...,\Delta_s\}$, whose entries are taken to be real but may be non-uniform.

For simplicity, we consider the case of a single isolated band at $T=0$, with the gap equation given by Eq. \eqref{eq:gapsingle}. The $\bd{q}=0$ gap equation reads
\begin{align}\label{eq:gapsingleq0}
\Delta_{\alpha}=&\frac{U_\alpha}{N}\sum_{\bd{k}}|u_{m\bd{k}\alpha}|^2\frac{\Delta_{m,\bd{k}}(0)}{2E_{m\bd{k}}},\,\,\,\,\,\,1\leqslant\alpha\leqslant s.
\end{align}
Here $\Delta_{m,\bd{k}}(0)=\la u_{m\bd{k}}|\hat{\Delta}|u_{m\bd{k}}\ra$ is the band-projected gap at $\bd{q}=0$, and $E_{m\bd{k}}=\sqrt{\xi_{m\bd{k}}^2+\Delta_{m,\bd{k}}(0)^2}$ is the quasiparticle energy.

Following the discussions at the end of App. \ref{app:derofgap} and computing the derivative of Eq. \eqref{eq:gapsingle}, we find
\begin{align}
M_{\alpha\beta}=\frac{\delta_{\alpha\beta}}{\Delta_\alpha}\sum_\bd{k}\frac{\Delta_{m,\bd{k}}(0)}{2E_{m\bd{k}}}|u_{m\bd{k}\alpha}|^2 -\sum_\bd{k}\frac{1}{2E_{m\bd{k}}}|u_{m\bd{k}\alpha}|^2|u_{m\bd{k}\beta}|^2,
\end{align}
and
\begin{align}
V_{\alpha,\mu}=\sum_\bd{k}\frac{1}{2E_{m\bd{k}}}[&(u_{m\bd{k}\alpha}^*\partial_\mu u_{m\bd{k}\alpha}-\partial_\mu u_{m\bd{k}\alpha}^* u_{m\bd{k}\alpha})\Delta_{m,\bd{k}}(0) \nonumber \\
&-2|u_{m\bd{k}\alpha}|^2\im \la u_{m\bd{k}}|\hat{\Delta}|\partial_\mu u_{m\bd{k}}\ra].
\end{align}
Now $U(1)$ symmetry imposes $\sum_{\beta=1}^sM_{\alpha\beta}\Delta_\beta=0$ and $\sum_{\beta=1}^sV_{\alpha,\mu}\Delta_\alpha=0$, so the kernel vector is instead $\bd{v}_0=(\Delta_1,\Delta_2,...,\Delta_s)^T$.

For the most general case, $\text{Rank}(M)=s-1$. To study the semidefiniteness, we take two-orbital models as an example. Note that the off-diagonal elements of $M_{\alpha\beta}$ are all negative, so we write
\begin{align}
M_{\alpha\beta}=\begin{pmatrix}
a & -b \\
-b & c
\end{pmatrix},
\end{align}
with $b>0$. Kernel eigenvector $\bd{v}_0$ leads to $a=(\Delta_2/\Delta_1)b$ and $c=(\Delta_1/\Delta_2)b$. Since $a,c$ are the diagonal entries of $M$, we conclude that for the two-orbital case, the Hessian matrix is positive semidefinite if and only if $\sgn(\Delta_1\Delta_2)=1$. This result largely but not completely coincides with the pair density-wave transition discovered in Ref. \cite{jiang2023} when one Hubbard interaction turns repulsive, $U_1U_2<0$. However, as a reminder, there it was also found that a solution channel with $\sgn(\Delta_1\Delta_2)=-1$ exists when both interactions are attractive, $U_1, U_2>0$.

For models with more than two orbitals, similar but more complicated conditions will be required for a nonpositive definite Hessian matrix.


\begin{thebibliography}{78}
\bibitem{footnote1}
The concept of interaction window is well-defined self-consistently. Starting from Hubbard models, the window can be first defined as the range of $2|U|$ around the Fermi level, with $U$ the largest Hubbard parameter. By solving the self-consistency gap equation and restricting to a channel, the order parameter $\Delta$ is obtained, which is typically smaller than $|U|$, so the window can be refined to $2\Delta$. After this step, some bands falling between $ \Delta $ and $|U|$ may be neglected. For pairing channels that are not uniform, matrix $\hat{\Delta}$ is not proportional to identity, then the window can be defined as $2\text{max}\{|\hat{\Delta}_{\alpha\beta}|\}$.
\bibitem{footnote2}
Two EBRs may share some atomic orbital components, in which case, since the inter-band quantum metric is negative semidefinite, it will still reduce the superfluid weight. However, this reduction is not substantial, as it is unable to change the lower bound by an integer.
\bibitem{footnote3}
This geometric transformation can be understood as physical, which is to shift the actual position of atomic orbitals but keep the hopping integrals fixed, or as artificial, which is to impose a componentwise phase for the Bloch function.
\bibitem{footnote4}
This projection is not orthogonal as the resulting vector is not orthogonal to $\bd{v}_0=(1,1,...,1)^T$. An orthogonal projection is not necessary since the purpose is only to get a simple expression of $D_s^{(2)}$.
\end{thebibliography}


%

\end{document}